\documentclass[authoryear]{elsarticle}

\usepackage{hyperref}
\usepackage{lineno}
\modulolinenumbers[5]
\usepackage{bm}
\usepackage{amsmath,amssymb}
\usepackage{subfigure}
\usepackage{color}
\usepackage{graphicx}

\graphicspath{{./}}

\journal{Icarus}

\begin{document}

\begin{frontmatter}

  \title{SPH Simulations for Shape Deformation of Rubble-Pile Asteroids Through Spinup: The Challenge for Making Top-Shaped Asteroids Ryugu and Bennu}

\author[mymainaddress]{Keisuke Sugiura\corref{mycorrespondingauthor}}
\cortext[mycorrespondingauthor]{Corresponding author}
\ead{sugiuraks@elsi.jp}

\author[myaddress3]{Hiroshi Kobayashi}

\author[myaddress2,myaddress4]{Sei-ichiro Watanabe}

\author[mymainaddress]{Hidenori Genda}

\author[myaddress4]{Ryuki Hyodo}

\author[myaddress3]{Shu-ichiro Inutsuka}

\address[mymainaddress]{Earth-Life Science Institute, Tokyo Institute of Technology, Tokyo 152-8550, Japan}

\address[myaddress3]{Department of Physics, Nagoya University, Aichi 464-8602, Japan}

\address[myaddress2]{Department of Earth and Environmental Sciences, Nagoya University, Aichi 464-8601, Japan}

\address[myaddress4]{Department of Solar System Sciences, ISAS, JAXA, Kanagawa 252-5210, Japan}

\begin{abstract}
  Asteroid 162173 Ryugu and asteroid 101955 Bennu, which were recently visited by spacecraft Hayabusa2 and OSIRIS-REx, respectively, are spinning top-shaped rubble piles. Other axisymmetric top-shaped near-Earth asteroids have been observed with ground-based radar, most of which rotate near breakup rotation periods of $\sim 3$ hours. This suggests that rotation-induced deformation of asteroids through rotational spinup produces top shapes. Although some previous simulations using the Discrete Element Method showed that spinup of rubble piles may produce oblate top shapes, it is still unclear what kinds of conditions such as friction angles of constituent materials and spinup timescales are required for top-shape formation. Here we show, through Smoothed Particle Hydrodynamics simulations of granular bodies spinning-up at different rates, that the rotation-induced deformation of spherical rubble piles before breakup can be classified into three modes according to the friction angle $\phi_{d}$: quasi-static and internal deformation for $\phi_{d} \leq 40^{\circ}$, dynamical and internal deformation for $50^{\circ} \leq \phi_{d} \leq 60^{\circ}$, and a set of surface landslides for $\phi_{d} \geq 70^{\circ}$. Note that these apparent large values of friction angle can be acceptable if we consider the effect of cohesion among blocks of a rubble pile under weak gravity. Bodies with $\phi_{d} \leq 60^{\circ}$ evolve into oblate spheroids through internal deformation, but never form pronounced equators defining a top shape. In contrast, bodies with $\phi_{d} \geq 70^{\circ}$ deform into axisymmetric top shapes through an axisymmetric set of surface landslides if spinup timescales are $\lesssim$ a few days. In addition, through slow spinups with timescales $\gtrsim 1$ month, bodies with $\phi_{d} \geq 70^{\circ}$ deform into non-axisymmetric shapes via localized sets of landslides. We suggest that rapid spinup mechanisms are preferable for the formation of axisymmetric top shapes.
\end{abstract}

\begin{keyword}
Asteroids, rotation; Asteroids, surfaces; Rotational dynamics; Regoliths 
\end{keyword}

\end{frontmatter}


\section{Introduction \label{Introduction}}
The Japan Aerospace Exploration Agency (JAXA) spacecraft Hayabusa2 arrived at asteroid 162173 Ryugu in June 2018 and investigated its detailed physical and geometrical properties via remote-sensing instruments. The proximity observation revealed that Ryugu is a rubble-pile asteroid with a so-called ``spinning-top'' shape having a prominent equatorial ridge (\citealt{Watanabe-et-al2019}). Radar observations already found some top-shaped near-Earth asteroids such as 1999 KW4, 2001 SN263, and 65803 Didymos (\citealt{Ostro-et-al2006, Becker-et-al2015, Benner-et-al2010}), and the recent proximity observation with the NASA spacecraft OSIRIS-REx confirmed that 101955 Bennu also has a top shape (\citealt{Lauretta-et-al2019, Barnouin-et-al2019}). Top-shaped asteroids have several common geometrical characteristics: polar-to-equatorial axis ratios of $\gtrsim 0.8$, pronounced equatorial ridges, nearly conical surfaces extending from the equators to the mid-latitudes, and almost axisymmetric shapes.

The fast rotation of a rubble-pile asteroid results in its deformation or breakup because the centrifugal force exceeds self-gravity at the equatorial radius, and the critical rotation period at which the centrifugal force equals self-gravity at the equator is $2 \pi / \sqrt{(4/3)\pi G \rho} \approx 3.3\,{\rm h}$ using a typical bulk density of C-type asteroids $\rho \approx 1 \,{\rm g/cm^{3}}$ (\citealt{Carry2012}), where $G$ is the gravitational constant. The spin periods of the top-shaped asteroids 1999 KW4, 2001 SN263, Didymos, and Bennu are $2.8\,{\rm h}$, $3.4\,{\rm h}$, $2.3\,{\rm h}$, and $4.3\,{\rm h}$, respectively (\citealt{Ostro-et-al2006, Becker-et-al2015, Richardson-et-al2016, Lauretta-et-al2019}). Although the current rotation period of Ryugu is $7.6\,{\rm h}$ (\citealt{Watanabe-et-al2019}), surface slope distribution and failure analysis of Ryugu suggest that Ryugu once had a rotation period of $\sim 3.5\,{\rm h}$ (\citealt{Watanabe-et-al2019, Hirabayashi-et-al2019}). This strongly suggests that Ryugu was reshaped by centrifugally induced deformation during the period of rapid rotation (\citealt{Watanabe-et-al2019}). Thus, other top-shaped asteroids may have also formed through deformation due to rapid rotation.

All the top-shaped asteroids found so far are near-Earth asteroids and have diameters $\lesssim$ several kilometers (\citealt{Ostro-et-al2006, Benner-et-al2010, Brozovic-et-al2011, Busch-et-al2011, Rozitis-et-al2014, Becker-et-al2015, Lauretta-et-al2019, Watanabe-et-al2019}). Their top shapes were identified through radar or proximity observations, which are mainly possible only for near-Earth asteroids. About half of the top-shaped asteroids have satellites (2001 NE263:\citealt{Becker-et-al2015}, 1999 KW4:\citealt{Ostro-et-al2006}, 1994 CC:\citealt{Brozovic-et-al2011}, Didymos:\citealt{Benner-et-al2010}). The orbits and sizes of the satellites and the fraction of the top-shaped asteroids with satellites may further constrain the formation process of top shapes.

Torque induced by the absorption of sunlight and its thermal re-emission changes the rotation of asteroids. The spinup or spindown mechanism is called the Yarkovsky-O'Keefe-Radzievskii-Paddack (YORP) effect (e.g.,\,\citealt{Rubincam2000, Walsh2018}), and the YORP spinup of several asteroids has already been confirmed (\citealt{Taylor-et-al2007, Kaasalainen-et-al2007, Hergenrother-et-al2019}). On the other hand, impacts with other small asteroids also change rotation states. Merging collisions and accretion of fragments after catastrophic collisions tend to increase spin rates due to angular momentum supply (e.g.,\,\citealt{Sugiura-et-al2018, Sugiura-et-al2019PSS, Michel-et-al2020}). The spinup of asteroids caused by such mechanisms eventually leads to global deformation, and sometimes probably results in top shapes (\citealt{Walsh-et-al2008}). The spinup also induces breakup of bodies to form satellites, which is consistent with the omnipresence of satellites around top-shaped asteroids.

How rapidly rotating rubble piles deform has been investigated in various studies through analytical approaches (\citealt{Holsapple2001, Holsapple2004, Sharma-et-al2009, Holsapple2010, Sharma2013, Sharma2018, Hirabayashi2015}) and numerical simulations (\citealt{Walsh-et-al2008, Walsh-et-al2012, Hirabayashi-et-al2015a, Sanchez-and-Scheeres2016, Zhang-et-al2017, Zhang-et-al2018, Sanchez-and-Scheeres2018, Leisner-et-al2020}). Most of the previous simulations about spinup of bodies have been conducted using the Discrete Element Methods (DEMs), in which usually spherical particles subject to gravity and contact forces represent the rubble-pile constituents. The simulations of spinup show that a body with a closely packed particle configuration experiences surface landslides and deforms into a top shape, while a body with a loosely packed particle configuration experiences internal deformation (\citealt{Walsh-et-al2008, Zhang-et-al2017}). This suggests that the interlocking of particles may affect deformation modes. DEM simulations correctly represent bulk friction of rubble piles if the sizes of DEM particles are comparable to those of blocks or regolith particles of asteroids. Practical sizes of DEM particles for asteroid simulations, however, are much larger than those of regolith particles due to computational limitations. The difference of the particle sizes possibly affects details of deformation modes of spinning-up bodies. Using a plastic finite element model (FEM) technique, some researches analyzed detailed relationship between cohesive strength and failure modes of rapidly rotating asteroids (e.g.,\,\citealt{Hirabayashi2015, Hirabayashi-et-al2015b, Hirabayashi-and-Scheeres2019}), while they could not investigate resultant shapes produced through the failure due to the limitation of FEM. Although the axisymmetric top-shaped figures are shown to be in equilibrium as a granular body (\citealt{Harris-et-al2009}), the formation process for such bodies is still unclear. Thus, the quantitative conditions of friction angle and acceleration rate required to form top shapes are still unclear.

The Smoothed Particle Hydrodynamics (SPH) method was recently applied to numerical simulations of shape deformation of asteroids (\citealt{Jutzi2015}). In the SPH simulations, a rubble-pile body is considered as a continuum of granular material and we explicitly set a specific value of friction angle of the material. Thus, we expect that the SPH method has the advantage of precise control of bulk friction of rubble piles. So far the SPH method has mainly been used to investigate shapes of impact outcomes (e.g.,\,\citealt{Jutzi-and-Asphaug2015, Jutzi-and-Benz2017, Leleu-et-al2018, Sugiura-et-al2018, Sugiura-et-al2019, Sugiura-et-al2019PSS}), and this is the first work applying the SPH method to the rotational deformation of small bodies.

In this paper, we present the results of SPH simulations about spinups of spherical rubble-pile bodies with various values of friction angles of constituent material. We then examine the friction angles required for the formation of top shapes. To clarify the spinup rate dependence, we also present the results of simulations with a slower spinup rate. Although the adopted rate is even much faster than the typical YORP spinup rate owing to the limitation of computational time, the comparison of these simulations with different spinup rates will help us to consider what kinds of acceleration mechanisms are appropriate for the top-shape formation.

\section{Method, simulation setup, and analysis of results \label{Method-simulation-setup-and-analysis-of-results}}
\subsection{Method \label{Method}}
The numerical code utilized in this study is based on a version of the SPH code described in \cite{Sugiura-et-al2018, Sugiura-et-al2019}. The code solves hydrodynamic equations (e.g.,\,the equations of continuity and motion) with self-gravity and yield processes with the aid of a friction model that determines shear strength of granular material (\citealt{Jutzi2015}). For the surface of a granular body, the time evolution of the density distribution of the body is calculated by solving the equation of continuity. The code is parallelized for distributed memory computers by using the Framework for Developing Particle Simulator (\citealt{Iwasawa-et-al2015, Iwasawa-et-al2016}). \cite{Sugiura2020} verified the validity of our SPH approach for granular materials through the comparison with laboratory experiments of cliff collapse (\citealt{Lajeunesse-et-al2005}).

We investigated the rotational deformation of a kilometer-sized granular body, where compressional stresses derived from self-gravity are too small for granular material to experience plastic compression. This enables us to use the following elastic equation of state:

\begin{equation}
  p = C_{s}^{2}(\rho - \rho_{0}),
  \label{elastic-EoS}
\end{equation}

\noindent where $p$ is the pressure, $\rho$ is the bulk density, $C_{s}$ is the bulk sound speed, and $\rho_{0}$ is the bulk density at uncompressed states. We set $\rho_{0}=1.19\,{\rm g/cm^{3}}$, which is the common value of the measured bulk densities of Ryugu and Bennu (\citealt{Watanabe-et-al2019, Lauretta-et-al2019}). We used a typical sound speed of granular material $\sim 100\,{\rm m/s}$ for $C_{s}$ (\citealt{Teramoto-and-Yano2005}). Simulation results only weakly depend on the adopted sound speed or the equations of state (see Appendix A). Note that we ignored the direct effect of cohesion, i.e., tensile forces, and thus we set $p=0$ if $p$ becomes negative.

\subsection{Simulation setup and analysis method \label{Simulation-setup-and-analysis-of-results}}
We mainly investigated the rotational deformation of a uniform and spherical body with a radius of $500\,{\rm m}$, which is similar to Ryugu's equatorial radius (\citealt{Watanabe-et-al2019}). Note that the deformation process and resultant shape are almost independent of the size of the body in our simulations (see Appendix A). We briefly mention the dependence of initial shapes on the deformation process by simulating bodies slightly different from the sphere in Section \ref{Spinup-of-bodies-with-equatorial-ridges}. The total number of SPH particles comprising the body is about 25,000 (or $\approx 20$ SPH particles in the radial direction), which has a sufficient resolution to capture shape-related features of a resultant body. The SPH particles were initially not placed on a regular lattice, but distributed randomly around the lattice points with a uniform density based on a procedure described by \cite{Sugiura-et-al2018}.

In general, the shear strength $Y_{d}$ of cohesive granular material as a function of the confining pressure $p$ can be represented as $Y_{d}=\tan (\phi_{d,{\rm real}})p + c_{{\rm coh}}$, where $\phi_{d,{\rm real}}$ and $c_{{\rm coh}}$ are the ``real'' friction angle and the cohesive strength, respectively. For simplicity, we adopted the relation in a form $Y_{d}=\tan (\phi_{d})p$ with the ``effective'' friction angle $\phi_{d}$ by interpreting that the relation $\phi_{d}=\tan^{-1}(Y_{d}/p)$ would be valid for cohesive granular materials. The effective friction angle $\phi_{d}$ used in our simulations can be expected to mimic an aspect of cohesion. The cohesion of a subsurface layer of asteroid Ryugu is estimated to be up to $c_{{\rm coh}}=670\,{\rm Pa}$ from an artificial impact experiment done in the Hayabusa2 mission (\citealt{Arakawa-et-al2020}), whereas the central pressure of a spherical body with a radius of $R = 500\,{\rm m}$ and mean density of $\rho_{0} = 1.19\,{\rm g/cm^{3}}$ is only $p_{{\rm c}} = (2/3)\pi G \rho_{0}^{2} R^{2} \approx 50\,{\rm Pa}$, so that the effective friction angle is up to $\phi_{d} \approx \tan^{-1}(c_{{\rm coh}}/p_{{\rm c}}) \approx 86^{\circ}$. To mimic large effective friction angles due to cohesion, we varied $\phi_{d}$ from $20^{\circ}$ to $80^{\circ}$. This treatment allows us to understand the detailed dependence of deformation modes on one parameter $\phi_{d}$. The validity of using such large effective friction angles instead of introducing the cohesion $c_{{\rm coh}}$ is discussed in Section \ref{Cohesion-and-large-effective-friction-angle}. 

We adopted an inertial coordinate system with origin at the center of mass of the body and the $z$-axis directed along the initial rotation axis of the body. The initial spherical shape of the body is stable for slow rotations, and the minimum rotation period $P_{{\rm min}}$ required for keeping the initial configuration gets shorter for larger $\phi_{d}$ (\citealt{Holsapple2001}). We set the initial rotation period of the body in each simulation with an effective friction angle $\phi_{d}$ to be longer than $P_{{\rm min}}$ for the corresponding $\phi_{d}$. To represent the spinup process of a body, we accelerated the rotation of the body around the $z$-axis as follows. We added an increment of velocity $\Delta \bm{v}_{i}$ in the rotational direction to the $i$-th SPH particle comprising the body at each numerical step as 

\begin{equation}
  \Delta \bm{v}_{i} = \Bigl( \dot{\omega}_{0} \Delta t \Bigr) \bm{e}_{z} \times \bm{r}_{i},
  \label{rotational-acceleration-of-main-body}
\end{equation}

\noindent where $\dot{\omega}_{0}$ is the predetermined angular acceleration of rotation, $\Delta t \approx 0.1\,{\rm s}$ is the time step, $\bm{e}_{z}$ is the unit vector parallel to the $z$-axis, and $\bm{r}_{i}$ is the position vector of the $i$-th SPH particle. The incremental velocity would increase the rotation rate by $\dot{\omega}_{0} \Delta t$ if its shape did not change. However, deformation actually changes the moment of inertia of the body, which results in a further change of the rotation rate. We set $\dot{\omega}_{0} = \beta \dot{\omega}_{n}$, where $\dot{\omega}_{n}=8.954\times 10^{-10}\,{\rm rad/s^{2}}$ is the angular acceleration in the nominal case defining the dimensionless parameter $\beta$. Here, we define the spinup timescale of a body as the elapsed time of spinup from the rotation period of $3.5\,{\rm h}$ to $3.0\,{\rm h}$. In the nominal case ($\beta=1$), the spinup timescale is $9.3 \times 10^{4}\,{\rm s}$, which corresponds to 8.0 rotations. The nominal rotational acceleration of $\beta=1$ is much faster than that due to the actual YORP effect for a kilometer-sized asteroid, which corresponds to $\beta \sim 10^{-8}$ (\citealt{Kaasalainen-et-al2007, Hergenrother-et-al2019}). We present the results of rotational deformation through spinups with $\beta=1$ in Section \ref{spinup-with-beta=1} and with much smaller $\beta$ in Section \ref{Spinup-with-beta=0.05}.

The spinup of the body eventually results in structural deformation and the ejection of surface masses. We accelerated only the SPH particles that comprise the main body, i.e., the largest clump of SPH particles, each of which has a distance from the nearest member of less than 1.5 times the smoothing length of a SPH particle. We identified clumps using a friends-of-friends algorithm (e.g.,\,\citealt{Huchra-and-Geller1982}).

Major deformation of the initial spherical body is followed by or accompanied by major mass ejection. In this study, we focused on the analysis of the shape of the main body formed through the major deformation event. To check the stability of the final shape of a deformed body, we stopped acceleration after 1\% of the total mass was ejected and continued the simulations for more than $1.0 \times 10^{5}\,{\rm s}$, which is longer than a typical deformation timescale of the body.

The ejected mass at some stage of a simulation is defined as the total mass of SPH particles that do not belong to the main body at that time. We measured the ratio $c/a$ of the minor to major axis of the main body using the top-down method (e.g.,\,\citealt{Capaccioni-et-al1984}), where the two axis lengths $a$ or $c$ were measured as the distances between two parallel plates that bound the main body. The rotation period $P$ of the main body is calculated as $P=2\pi I_{z}/L_{z}$, where $I_{z}$ and $L_{z}$ are, respectively, the moment of inertia and the angular momentum of the main body around the $z$-axis. Note that the angle between the angular momentum vector and the $z$-axis is $< 0.3^{\circ}.$ Although we do not do any corrections for the rotation axis of the body, the rotation axis is almost aligned with the $z$-axis. Thus, $P$ calculated from $I_{z}$ and $L_{z}$ is a reliable measure for the rotation period.

\section{Results \label{Results}}

\subsection{Spinup with $\beta=1$ \label{spinup-with-beta=1}}

\begin{figure}[!htb]
 \begin{center}
 \includegraphics[bb=0 0 994 568, width=1.0\linewidth]{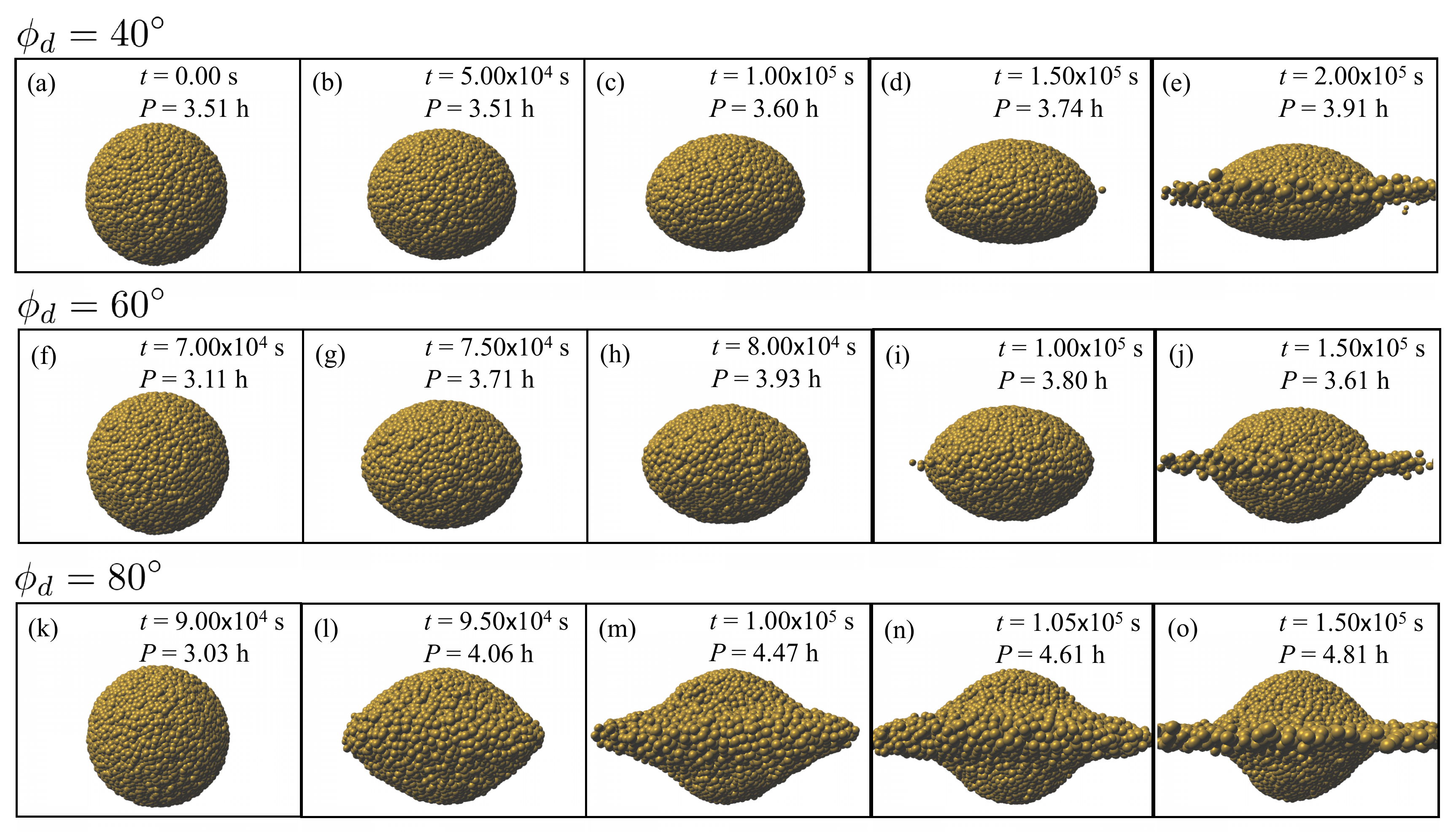}
 \caption{Side-view snapshots of spinning-up bodies with effective friction angles $\phi_{d}=40^{\circ}$ (a-e), $60^{\circ}$ (f-j), and $80^{\circ}$ (k-o). Each panel includes elapsed time $t$ and the instantaneous rotation period of the main body $P$. Note that $t$ for $\phi_{d}=60^{\circ}$ and $80^{\circ}$ is the elapsed time from the start of the simulations, while that for $\phi_{d}=40^{\circ}$ is from $2.0\times 10^{5}\,{\rm s}$ after the start of the simulation.}
 \label{pictures-gsr-R=500m-N=25000-nominalSlowAccel} 
 \end{center}
\end{figure}

\begin{figure}[!htb]
 \begin{center}
 \includegraphics[bb=0 0 370 252, width=1.0\linewidth]{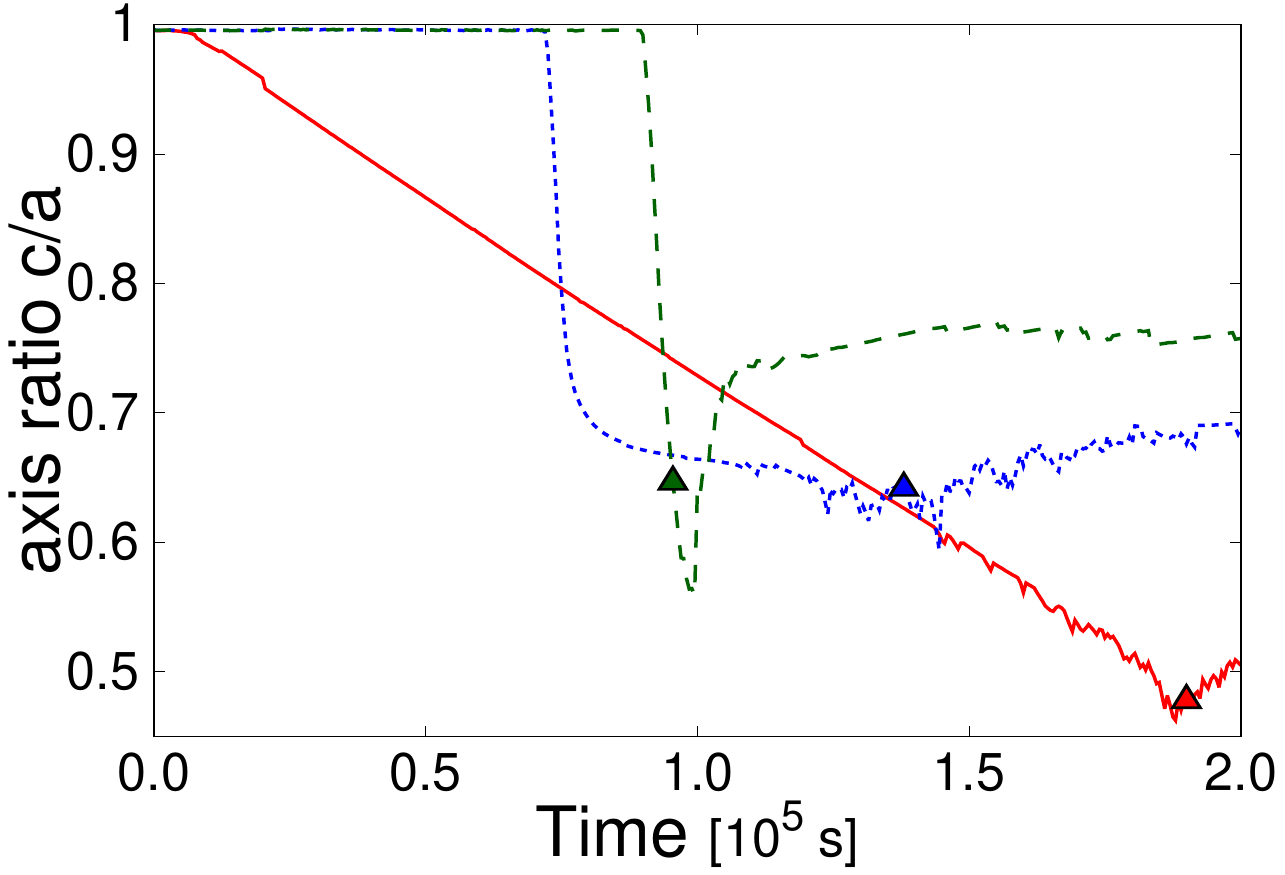}
 \caption{Time evolution of the axis ratios $c/a$ of the main body in the cases of $\phi_{d}=40^{\circ}$ (red solid curve), $\phi_{d}=60^{\circ}$ (blue dotted curve), and $\phi_{d}=80^{\circ}$ (green dashed curve). Note that time on the horizontal axis for $\phi_{d}=40^{\circ}$ shows the elapsed time from $2.0\times 10^{5}{\rm s}$. The filled triangles indicate $c/a$ at the times when the spinup is halted (when 1\% of the total mass is ejected).}
 \label{time-vs-c_a-R=500m-N=25000-phid=40_60_80-nominalSlowAccel} 
 \end{center}
\end{figure}

In simulations with the nominal spinup rate $\beta=1$, we set the initial rotation periods for $\phi_{d} \leq 40^{\circ}$ to be $5.5\,{\rm h}$ and those for $\phi_{d} \geq 50^{\circ}$ to be $3.5\,{\rm h}$ in order to save the computational time because in the latter cases deformation of the body never occurs during the periods when $P > 3.5\,{\rm h}$. Figure \ref{pictures-gsr-R=500m-N=25000-nominalSlowAccel} shows side-view snapshots of the $\beta = 1$ spinup of the granular spherical bodies with the effective friction angles $\phi_{d}=40^{\circ}$, $60^{\circ}$, and $80^{\circ}$. Figure \ref{time-vs-c_a-R=500m-N=25000-phid=40_60_80-nominalSlowAccel} shows a time evolution of the axis ratio $c/a$ of the main body obtained from the same simulations. Because of the difference of the initial rotation periods, time when deformation occurs for $\phi_{d}=40^{\circ}$ is $\approx 2.0 \times 10^{5}\,{\rm s}$ later than those for $\phi_{d}=60^{\circ}$ and $80^{\circ}$. For better comparison, elapsed time from $2.0 \times 10^{5}\,{\rm s}$ after the start of the simulation is shown for $\phi_{d}=40^{\circ}$ in Figs.\,\ref{pictures-gsr-R=500m-N=25000-nominalSlowAccel}a-e and \ref{time-vs-c_a-R=500m-N=25000-phid=40_60_80-nominalSlowAccel}.

All the deformed bodies in Fig.\,\ref{pictures-gsr-R=500m-N=25000-nominalSlowAccel} have kept almost axisymmetric shapes, indicating that the spinup of a spherical body with $\beta=1$ results in axisymmetric deformation. This behavior is quite different from a self-gravitating fluid body, for which rapid rotation induces a non-axisymmetric instability to form a series of Jacobi triaxial ellipsoids (\citealt{Chandrasekhar1969})\footnote{The stability of granular asteroids has been investigated by \cite{Sharma2013} in detail. Although the work is relevant to our simulations, direct comparison between the work and our simulations is difficult because setups are different. Especially, our present study simulates bodies in which global plastic deformation hardly occurs, so that analysis of \cite{Sharma2013} on plastic modulus is not expected to affect our present study. In future work, we are planning to perform simulations that use the same setup as the work and discuss application for the stability of granular asteroids.}. The non-axisymmetric deformation into triaxial ellipsoids with small intermediate-to-major axis ratio $< 0.5$ occurs only for $\phi_{d} \leq 20^{\circ}$. This suggests that relatively small values of $\phi_{d}$ ($\sim 30^{\circ}$) of granular bodies will prevent the deformation into triaxial ellipsoids. Note that non-axisymmetric sets of landslides occurring in slower spinup of bodies with $\phi_{d} \geq 70^{\circ}$ (see Section \ref{Spinup-with-beta=0.05}) belong to a different deformation mode from the internal deformation forming non-axisymmetric triaxial ellipsoids for $\phi_{d} \leq 20^{\circ}$.

For $\phi_{d}=40^{\circ}$, the axis ratio $c/a$ decreases from unity (Fig.\,\ref{pictures-gsr-R=500m-N=25000-nominalSlowAccel}a) to $\approx 0.5$ (Fig.\,\ref{pictures-gsr-R=500m-N=25000-nominalSlowAccel}d) during $1.0\times 10^{4}\,{\rm s} - 1.9\times 10^{5}\,{\rm s}$ (Fig.\,\ref{time-vs-c_a-R=500m-N=25000-phid=40_60_80-nominalSlowAccel}). Timescale of the deformation is comparable to the spinup timescale $\approx 9.3\times 10^{4}\,{\rm s}$, and the deformation of the body almost halts once we stop the spinup. We define this spinup-driven deformation mode as a quasi-static deformation. Further acceleration of the rotation during $t > 1.9\times 10^{5}\,{\rm s}$ leads to mass ejection from the body's surface in the equatorial plane (Figs.\,\ref{pictures-gsr-R=500m-N=25000-nominalSlowAccel}e and \ref{time-vs-c_a-R=500m-N=25000-phid=40_60_80-nominalSlowAccel}). Note that the body spins down because of the increase in the moment of inertia due to the deformation of the body, although the angular momentum increases owing to the imposed spinup condition. In addition, the vertically asymmetric shape is formed (see Figs.\,\ref{pictures-gsr-R=500m-N=25000-nominalSlowAccel}c-e), which is probably caused by initial asymmetry of SPH particle distribution along $z$-direction.

For $\phi_{d}=60^{\circ}$, the initial spherical shape remains unchanged until $t \approx 7 \times 10^{4}\,{\rm s}$ when a rapid deformation occurs (Fig.\,\ref{time-vs-c_a-R=500m-N=25000-phid=40_60_80-nominalSlowAccel}). The deformation timescale of $\sim 10^{4}\,{\rm s}$ is much shorter than the spinup timescale $\approx 9.3\times 10^{4}\,{\rm s}$. Once the significant deformation occurs, the deformation is promoted even without the spin acceleration. We define this deformation mode as a dynamical deformation. After $t \approx 7 \times 10^{4}\,{\rm s}$, the decrease rate of the axis ratio slows until $t \approx 1.4 \times 10^{5}\,{\rm s}$. Significant mass ejection occurs at $t \approx 1.4 \times 10^{5}\,{\rm s}$ (Fig.\,\ref{time-vs-c_a-R=500m-N=25000-phid=40_60_80-nominalSlowAccel}). The mass ejection under the slow-deformation phase seems similar to the case with $\phi_{d}=40^{\circ}$.

For $\phi_{d}=80^{\circ}$, the axis ratio suddenly decreases at $t \approx 9.0 \times 10^{4}\,{\rm s}$, which is a similar dynamical deformation to the case with $\phi_{d}=60^{\circ}$ (Fig.\,\ref{time-vs-c_a-R=500m-N=25000-phid=40_60_80-nominalSlowAccel}). However, the mass ejection from the surface in the equatorial plane is accompanied by the deformation of the main body (Figs.\,\ref{pictures-gsr-R=500m-N=25000-nominalSlowAccel}n and \ref{time-vs-c_a-R=500m-N=25000-phid=40_60_80-nominalSlowAccel}). Significant deformation begins at the time when $P=3.03\,{\rm h}$ (Fig.\,\ref{pictures-gsr-R=500m-N=25000-nominalSlowAccel}k). At the equator, the centrifugal forces are almost equal to the gravitational attraction of the body. Thus, the deformation pushes the material near the equator out of the surface of the body, which rapidly leads to mass ejection (Fig.\,\ref{pictures-gsr-R=500m-N=25000-nominalSlowAccel}n). The amount of the ejected mass is $\sim 10\%$ of the main body. The angular momentum loss due to the mass ejection increases the rotation period, which stalls further deformation. The main body finally has axis ratio $c/a = 0.76$ (Fig.\,\ref{time-vs-c_a-R=500m-N=25000-phid=40_60_80-nominalSlowAccel}) and rotation period $P = 4.8\,{\rm h}$ (Fig.\,\ref{pictures-gsr-R=500m-N=25000-nominalSlowAccel}o). The final shape seems similar to a top shape (Fig.\,\ref{pictures-gsr-R=500m-N=25000-nominalSlowAccel}o, see also the supplementary movie), but the axis ratio $c/a = 0.76$ is smaller than that of Ryugu $c/a = 0.87$ and Bennu $c/a = 0.88$ (\citealt{Watanabe-et-al2019, Lauretta-et-al2019}).

\begin{figure}[!htb]
 \begin{center}
 \includegraphics[bb=0 0 1701 595, width=1.0\linewidth]{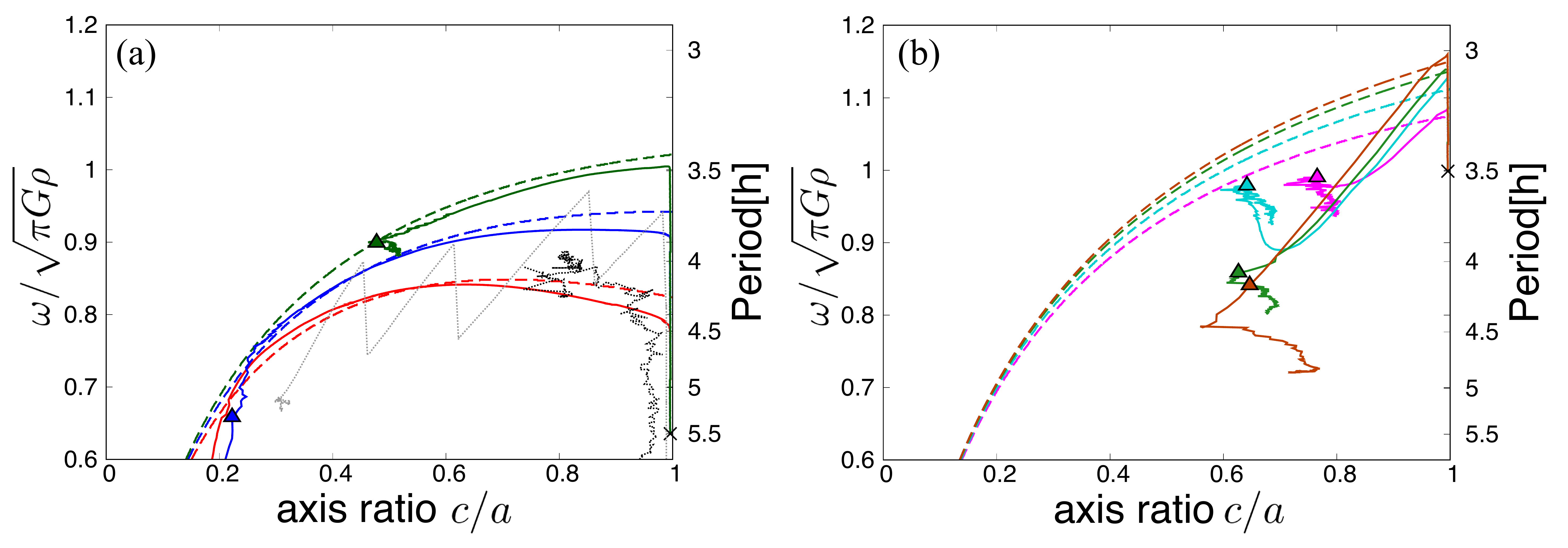}
 \caption{Evolutionary tracks of the main bodies in instantaneous axis ratio $c/a$ versus spin rate diagrams obtained from our simulations (solid curves) in comparison with the maximum equilibrium spins of oblate spheroids given by an analytic formula found by \cite{Holsapple2001} and \cite{Sharma-et-al2009} (dashed curves). The horizontal axis shows the ratio $c/a$ of the minor to major axis. The right and left vertical axes show, respectively, the rotation period for $\rho = 1.19\,{\rm g/cm^{3}}$ and the angular speed $\omega$ scaled by $\sqrt{\pi G \rho}$, where $G$ is the gravitational constant. (a) Red, blue, and green curves show the cases for the effective friction angle $\phi_{d}=20^{\circ}$, $30^{\circ}$, and $40^{\circ}$, respectively. (b) Magenta, cyan, light green, and brown curves show the cases for $\phi_{d}=50^{\circ}$, $60^{\circ}$, $70^{\circ}$, and $80^{\circ}$, respectively. For comparison with the low $\phi_{d}$ case shown on the panel (a), we plot evolutionary tracks of axis ratio and normalize spin rate obtained from two DEM spinup simulations: the black dotted curve is an HSDEM model from \cite{Walsh-et-al2012}; the gray dotted curve is an SSDEM model from \cite{Sanchez-and-Scheeres2012}. The crosses and filled triangles on the solid curves show the initial states and the states at the times when the spinup is halted (when 1\% of the total mass is ejected), respectively. For $\phi_{d}=20^{\circ}$, the mass ejection occurs when $P = 7.6\,{\rm h}$ and $c/a = 0.16$, so that the red triangle is out of the panel (a).}
 \label{comparison-gsr-R=500m-N=25000-nominalSlowAccel-with-Holsapple2001-c_a-vs-scaledOmega} 
 \end{center}
\end{figure}

Figure \ref{comparison-gsr-R=500m-N=25000-nominalSlowAccel-with-Holsapple2001-c_a-vs-scaledOmega} shows the instantaneous spin rate $\omega$ of the main body as a function of axis ratio $c/a$ obtained from our numerical simulations with various effective friction angles in comparison with the maximum equilibrium spin of oblate spheroids given by an analytic formula found by \cite{Holsapple2001} and \cite{Sharma-et-al2009}. For the cases with low effective friction angles $\phi_{d}=20^{\circ}$, $30^{\circ}$, and $40^{\circ}$, $\omega$ initially increases while the bodies maintain their spherical shapes ($c/a=1$) until $\omega$ approaches the maximum equilibrium angular speed. Then $\omega$ and $c/a$ vary mostly according to the maximum equilibrium spin state. The mass ejection begins after significant deformation with $c/a < 0.5$.

Figure \ref{comparison-gsr-R=500m-N=25000-nominalSlowAccel-with-Holsapple2001-c_a-vs-scaledOmega}a also shows the normalized spin rates and axis ratios of spinning-up bodies simulated using the DEMs for cohesionless granular materials: the Hard-Sphere DEM (HSDEM) simulation in \cite{Walsh-et-al2012} (red solid curve in Fig.\,9 of the paper) and the Soft-Sphere DEM (SSDEM) simulation in \cite{Sanchez-and-Scheeres2012} (red solid curve in Fig.\,26 of the paper). We only plot the data of the DEM simulations until major mass ejection events. The precise evaluation of the bulk friction is difficult for the DEM simulations, but the DEM results roughly correspond to the maximum equilibrium spin curves with $\phi_{d}=20 - 30^{\circ}$. Although not only the methods but also the particle numbers, spinup rates, and spinup procedures are different from those of our simulations, the evolution paths of the DEM simulations are consistent with those of our simulations with $\phi_{d} = 20 - 30^{\circ}$. On the other hand, the axis ratio $c/a$ of the HSDEM simulation at the epoch of mass ejection is largely different from those of the SSDEM and our simulations: $c/a \approx 0.8$, $0.3$, and $0.2$ for the HSDEM, the SSDEM, and our simulation with $\phi_{d}=30^{\circ}$, respectively. This suggests that SSDEM and SPH simulations provide more similar results than HSDEM simulations.

For $\phi_{d}=50^{\circ}$, $60^{\circ}$, $70^{\circ}$, and $80^{\circ}$, $\omega$ exceeds the maximum equilibrium angular speed without deformation followed by sudden dynamical deformation in which both $c/a$ and $\omega$ decrease (Fig.\,\ref{comparison-gsr-R=500m-N=25000-nominalSlowAccel-with-Holsapple2001-c_a-vs-scaledOmega}b). These are quite different from the quasi-equilibrium evolution path for lower $\phi_{d}$ (Fig.\,\ref{comparison-gsr-R=500m-N=25000-nominalSlowAccel-with-Holsapple2001-c_a-vs-scaledOmega}a). The dynamical deformation timescale is much shorter than the spinup timescale (see Fig.\,\ref{time-vs-c_a-R=500m-N=25000-phid=40_60_80-nominalSlowAccel}). The sudden deformation allows the spin rate to be smaller than the maximum spin rate in equilibrium. Mass ejection occurs after the dynamical deformation for $\phi_{d}=50^{\circ}$ and $60^{\circ}$, while during the dynamical deformation for $\phi_{d}=70^{\circ}$ and $80^{\circ}$.

\begin{figure}[!htb]
 \begin{center}
 \includegraphics[bb=0 0 960 785, width=1.0\linewidth]{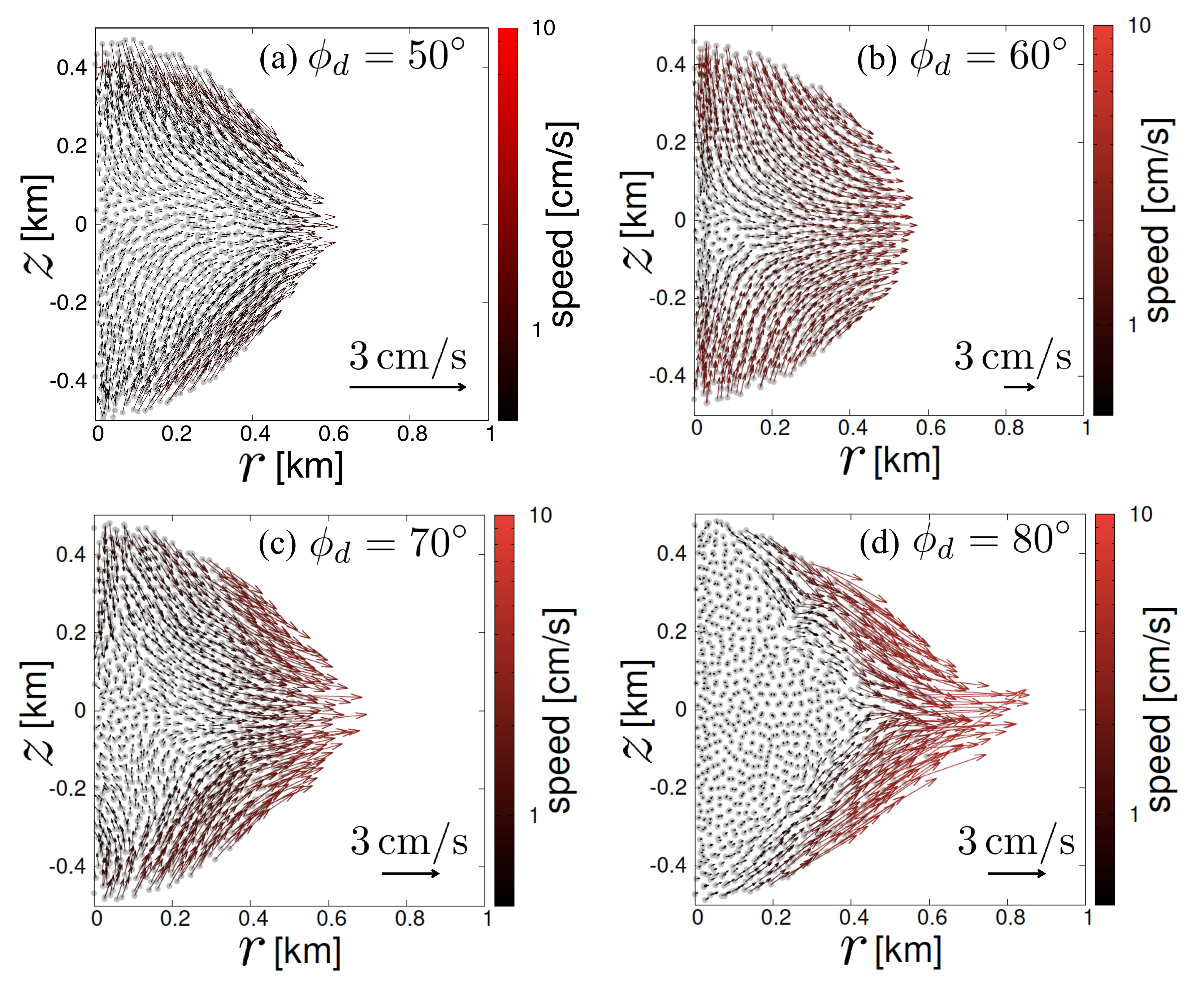}
 \caption{Meridional cross-sections of the main bodies with the effective friction angles $\phi_{d} = 50^{\circ}$ (a), $60^{\circ}$ (b), $70^{\circ}$ (c), and $80^{\circ}$ (d) during the dynamical deformation. These snapshots are taken at $5.0\times 10^{4}\,{\rm s}$ (a), $7.4\times 10^{4}\,{\rm s}$ (b), $8.0\times 10^{4}\,{\rm s}$ (c), and $9.3\times 10^{4}\,{\rm s}$ (d) after the start of each simulation. The horizontal axis shows the distance from the rotation axes $r$, and the vertical axis shows the height from the equatorial planes $z$. The arrows and colors show the velocity vectors and the speeds of the SPH particles in the planes of the cross-sections. Note that the deformation occurs axisymmetrically (see Fig.\,\ref{pictures-gsr-R=500m-N=25000-nominalSlowAccel}).}
 \label{2DPositionAndVelocityAtDeformation-gsr-R=500m-N=25000-nominalSlowAccel} 
 \end{center}
\end{figure}

Figure \ref{2DPositionAndVelocityAtDeformation-gsr-R=500m-N=25000-nominalSlowAccel} shows the cross-sectional snapshots of the main bodies with different $\phi_{d}$ during the dynamical deformation. The simulations with $\phi_{d} = 50^{\circ}$ and $60^{\circ}$ (Fig.\,\ref{2DPositionAndVelocityAtDeformation-gsr-R=500m-N=25000-nominalSlowAccel}a and b) show that the deformation speeds at the polar regions and the equatorial regions are almost the same, and not only the surface but also the deep interior of the bodies deform. A typical flow pattern can be seen in the case with $\phi_{d} = 60^{\circ}$ (Fig.\,\ref{2DPositionAndVelocityAtDeformation-gsr-R=500m-N=25000-nominalSlowAccel}b), where streamlines are similar to rectangular hyperbolae that tend to flatten the body. We define this deformation mode as internal deformation.

For $\phi_{d} = 70^{\circ}$ and $80^{\circ}$ (Fig.\,\ref{2DPositionAndVelocityAtDeformation-gsr-R=500m-N=25000-nominalSlowAccel}c and d), deformation speeds in the surface layers from equator to mid-latitudes are faster than those in polar and internal regions, and deformation vectors in the surface layers point toward the equators. The typical flow pattern can be seen for $\phi_{d}=80^{\circ}$ (Fig.\,\ref{2DPositionAndVelocityAtDeformation-gsr-R=500m-N=25000-nominalSlowAccel}d), where deformation is negligible in the internal region and the deformation speeds are significant only at the surface layers. We define this deformation mode as surface landslides. The deformation occurs at the time when $P = 3.07\,{\rm h}$ for $\phi_{d}=70^{\circ}$ and $P = 3.03\,{\rm h}$ for $\phi_{d}=80^{\circ}$. These rotation periods are very close to the rotational breakup period $P_{{\rm lim}} = 2 \pi / \sqrt{(4/3)\pi G \rho_{0}} = 3.027\,{\rm h}$ for a spherical body without tensile strength. Thus, the deformation directly causes significant mass ejection (see also Fig.\,\ref{pictures-gsr-R=500m-N=25000-nominalSlowAccel}k-m), and the sudden mass ejection from the equatorial surface triggers a set of surface landslides. The mass ejection simultaneously occurs at various longitudes, which induces an axisymmetric occurrence of multiple landslides at almost the same time. We refer to the axisymmetric and simultaneous occurrence of multiple landslides as an axisymmetric set of landslides.

In summary, three deformation modes according to the effective friction angle $\phi_{d}$ are identified through our simulations with $\beta=1$: quasi-static and internal deformation for $\phi_{d} \leq 40^{\circ}$ (Fig.\,\ref{pictures-gsr-R=500m-N=25000-nominalSlowAccel}a-e and Fig\,\ref{comparison-gsr-R=500m-N=25000-nominalSlowAccel-with-Holsapple2001-c_a-vs-scaledOmega}a); dynamical and internal deformation for $50^{\circ} \leq \phi_{d} \leq 60^{\circ}$ (Fig.\,\ref{2DPositionAndVelocityAtDeformation-gsr-R=500m-N=25000-nominalSlowAccel}a and b); and an axisymmetric set of surface landslides for $ \phi_{d} \geq 70^{\circ}$ (Fig.\,\ref{2DPositionAndVelocityAtDeformation-gsr-R=500m-N=25000-nominalSlowAccel}c and d).

\begin{figure}[!htb]
 \begin{center}
 \includegraphics[bb=0 0 960 916, width=1.0\linewidth]{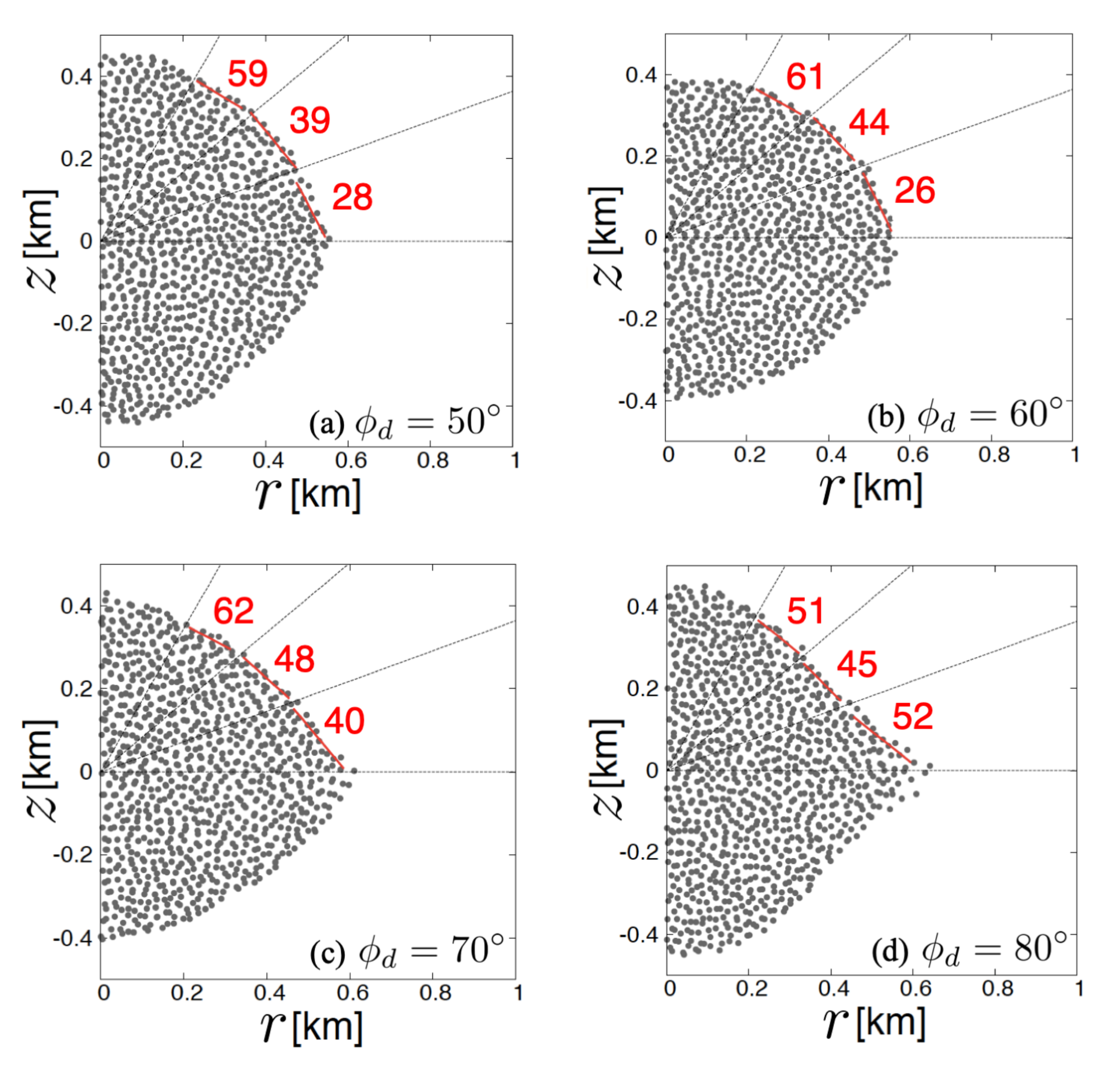}
 \caption{Meridional cross-sections of the main body with effective friction angles $\phi_{d} = 50^{\circ}$ (a), $60^{\circ}$ (b), $70^{\circ}$ (c), and $80^{\circ}$ (d) at $t=4.0\times 10^{5}\,{\rm s}$, sufficiently after the onset of mass ejection. In each panel, red numbers represent the surface tilt angles [degree], i.e., the angles between the rotation axes and the surfaces in the corresponding latitudinal ranges. The dotted lines show the latitudes of $0^{\circ}$, $20^{\circ}$, $40^{\circ}$, and $60^{\circ}$.}
 \label{2DPositionAtFinal-gsr-R=500m-N=25000-nominalSlowAccel} 
 \end{center}
\end{figure}

Figure \ref{2DPositionAtFinal-gsr-R=500m-N=25000-nominalSlowAccel} shows the cross-sections of the main bodies at the end of the simulations. For $\phi_{d}=50^{\circ}$ and $60^{\circ}$, oblate spheroidal shapes without equatorial ridges are formed (Fig.\,\ref{2DPositionAtFinal-gsr-R=500m-N=25000-nominalSlowAccel}a and b). The surface tilt angles relative to the $z$-direction are $\sim 30^{\circ}$ at low latitudes (latitudes of $0^{\circ} - 20^{\circ}$) and $\sim 60^{\circ}$ at mid latitudes (latitudes of $40^{\circ} - 60^{\circ}$). The large tilt angle differences of $\sim 30^{\circ}$ between the low- and mid-latitudes show that the main body has a near-spheroidal surface.

For $\phi_{d}=70^{\circ}$ and $80^{\circ}$, the final shapes of the main bodies seem to have conical surfaces rather than spheroidal surfaces (Fig.\,\ref{2DPositionAtFinal-gsr-R=500m-N=25000-nominalSlowAccel}c and d). The difference of the surface tilt angles between the low- and mid-latitudes is $\sim 20^{\circ}$ for $\phi_{d}=70^{\circ}$ and $\sim 7^{\circ}$ for $\phi_{d}=80^{\circ}$. Ryugu has differences of the surface tilt angles between low- and mid-latitudes $\lesssim 20^{\circ}$ at almost all longitudes (\citealt{Watanabe-et-al2019}). Thus the final shapes resulting from our simulations with $\phi_{d} \geq 70^{\circ}$ have conical surfaces similar to that of Ryugu. Such conical surfaces are most likely produced by landslides. Moreover, a top shape has a characteristic that the topographic distance from the center of the body has a minimum at a mid-latitude, and that characteristic is reproduced for the bodies formed through the simulations with $\phi_{d} \geq 70^{\circ}$. Therefore, top shapes can be produced through an axisymmetric set of landslides caused by spinup with $\beta = 1$ and $\phi_{d} \geq 70^{\circ}$ (Figs.\ref{pictures-gsr-R=500m-N=25000-nominalSlowAccel}o and \ref{2DPositionAndVelocityAtDeformation-gsr-R=500m-N=25000-nominalSlowAccel}d).

One cannot distinguish top shapes from spheroids by their axis ratios. Here, we adopt a quantitative method to distinguish top shapes from spherical shapes. We define the relative volume difference $\delta_{V}$ as

\begin{equation}
  \delta_{V} = (V_{e} - V)/V_{e},
  \label{difference-from-ellipsoidal-volume}
\end{equation}

\noindent where $V$ and $V_{e}$ are the volume of a body and an ellipsoid with the same triaxial dimensions as the body, respectively. We measured the axis length of the body by the top-down method (e.g.,\,\citealt{Capaccioni-et-al1984}). Bodies with ellipsoidal or rounded shapes have $\delta_{V} \ll 1$, while bodies with angular shapes such as top shapes have large $\delta_{V}$. Ryugu has $\delta_{V} \approx 0.3$ (\citealt{Watanabe-et-al2019}). 

\begin{figure}[!htb]
 \begin{center}
 \includegraphics[bb=0 0 720 526, width=1.0\linewidth]{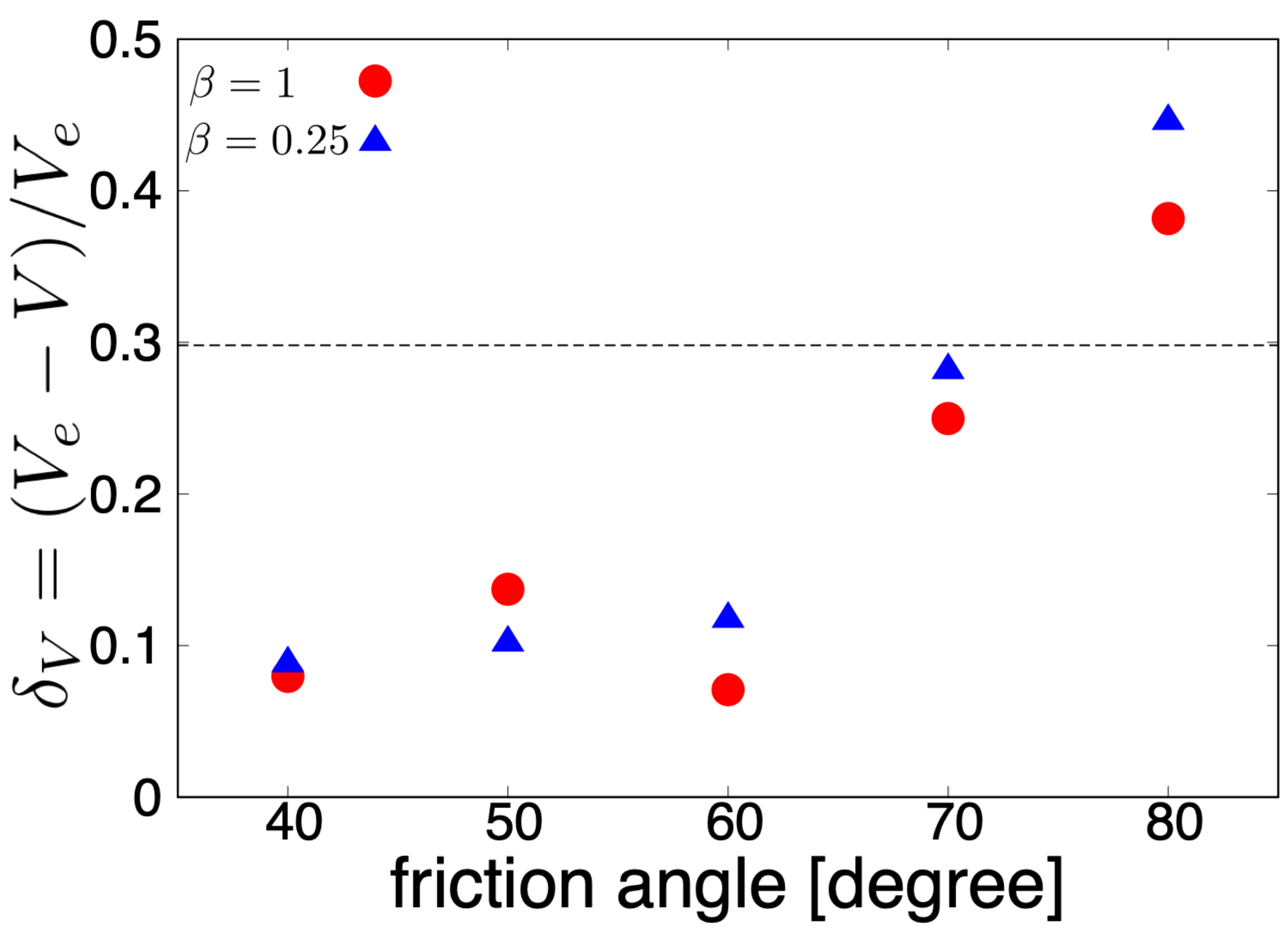}
 \caption{Difference $\delta_{V}$ from ellipsoidal volume of the main body produced through the $\beta=1$ (red circles) and $\beta=0.25$ (blue triangles) spinup of the spherical bodies with various effective friction angles. The horizontal axis shows the effective friction angle used in each simulation, and the vertical axis shows $\delta_{V}$ of the main body obtained from each simulation. The dotted line shows $\delta_{V}$ of asteroid Ryugu obtained from \cite{Watanabe-et-al2019}.}
 \label{comparisonWithElipsoidalVolumeWithTopDown-gsr-R=500m-rho=1.19-nominalSlowAccel} 
 \end{center}
\end{figure}

Figure \ref{comparisonWithElipsoidalVolumeWithTopDown-gsr-R=500m-rho=1.19-nominalSlowAccel} shows $\delta_{V}$ of the main body resulting from the spinup of the spherical body with $\beta=1$ and $0.25$. For both $\beta$ values, $\delta_{V}$ is $\approx 0.1$ for $\phi_{d}=40^{\circ}$, $50^{\circ}$, and $60^{\circ}$ and increases with increasing $\phi_{d}$ for $\phi_{d}>60^{\circ}$. For $\phi_{d}=70^{\circ}$ and $80^{\circ}$, $\delta_{V}$ is close to or larger than that of Ryugu. This tendency is consistent with the results of our simulations: the body has spheroidal shapes for $\phi_{d} \leq 60^{\circ}$ but angular shapes for $\phi_{d} \geq 70^{\circ}$ (Fig.\,\ref{2DPositionAtFinal-gsr-R=500m-N=25000-nominalSlowAccel}). Thus, we quantitatively confirmed that our simulations produce top shapes for $\phi_{d} \geq 70^{\circ}$.

It should be noted that the top shape produced in the simulation with $\phi_{d}=80^{\circ}$ has $c/a = 0.76$, while almost all of the top shapes found so far have $c/a > 0.85$ (e.g.,\,\citealt{Ostro-et-al2006, Busch-et-al2011, Watanabe-et-al2019, Lauretta-et-al2019}). Landslides mainly induce material transfer from high latitudes to equatorial regions (see Fig.\,\ref{2DPositionAndVelocityAtDeformation-gsr-R=500m-N=25000-nominalSlowAccel}d), and thus it is basically difficult for bodies to keep $c/a$ of about unity through landslides. The discrepancy regarding $c/a$ between the observed top shapes and that produced through our simulations should be considered in future work.

\subsection{Spinup with $\beta=0.05$ \label{Spinup-with-beta=0.05}}
The deformation processes of the body with $\phi_{d} \leq 60^{\circ}$ under spinup with $\beta=0.05$ are similar to those with $\beta=1$, i.e., the quasi-static, axisymmetrical, and internal deformation occurs for $\phi_{d}<40^{\circ}$, and the dynamical, axisymmetrical, and internal deformation occurs for $50^{\circ} \leq \phi_{d} \leq 60^{\circ}$. The resultant shapes are axisymmetric. Thus, the spinup rate $\beta$ does not seem to affect the deformation modes of the body with a low effective friction angle $\phi_{d} \leq 60^{\circ}$.

\begin{figure}[!htb]
 \begin{center}
 \includegraphics[bb=0 0 1332 710, width=1.0\linewidth]{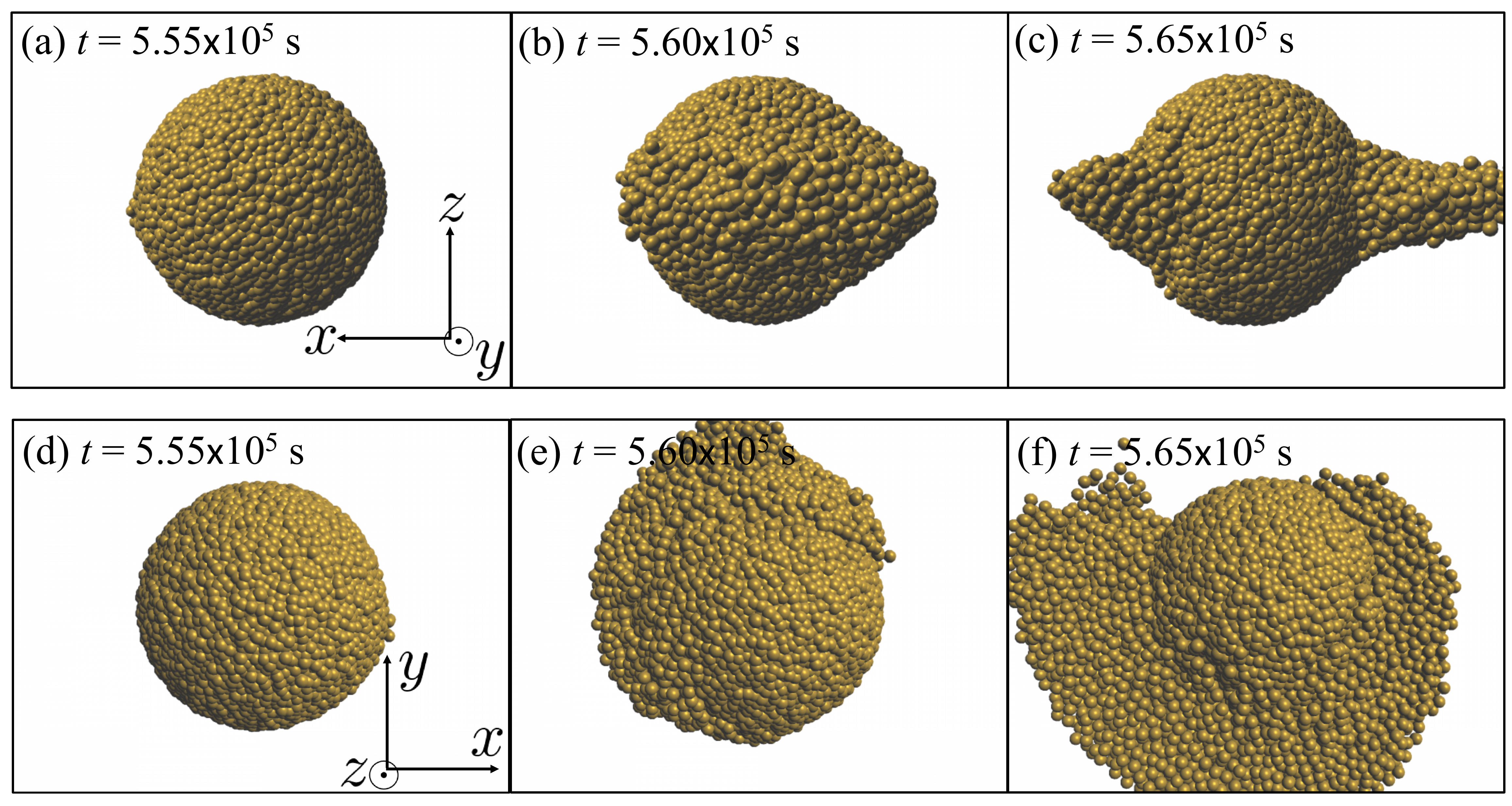}
 \caption{Snapshots of the body with $\phi_{d}=80^{\circ}$ and $\beta = 0.05$. Panels (a)-(c) show snapshots taken from the $+y$-direction, where the line of sight is perpendicular to the rotation axis. Panels (d)-(f) show snapshots taken from the $+z$-direction, where the line of sight is parallel to the rotation axis.}
 \label{pictures-gsr-R=500m-N=25000-0.05NominalSlowAccel} 
 \end{center}
\end{figure}

In contrast, simulations of the body with $\phi_{d} \geq 70^{\circ}$ under spinup with $\beta=0.05$ result in a different deformation mode from those under $\beta=1$. Figure \ref{pictures-gsr-R=500m-N=25000-0.05NominalSlowAccel} shows the snapshots of the simulation with $\phi_{d}=80^{\circ}$ and $\beta = 0.05$. The initial rotation period of the main body is set to be $3.25\,{\rm h}$. At $t \approx 5.60 \times 10^{5}\,{\rm s}$, the rotation period becomes $P \approx 3.08\,{\rm h}$ and the mass ejection mainly starts from an equatorial region (Fig.\,\ref{pictures-gsr-R=500m-N=25000-0.05NominalSlowAccel}b and e). At $t \approx 5.65 \times 10^{5}\,{\rm s}$, a significant amount of mass ($\sim 10\%$ of the total mass) is ejected, but it does not come from an equatorial region around the $+y$-direction (Fig.\,\ref{pictures-gsr-R=500m-N=25000-0.05NominalSlowAccel}c and f). Therefore, non-axisymmetric mass ejection occurs. The rotation period becomes $P \approx 4.5\,{\rm h}$ due to angular momentum loss through the mass ejection.

\begin{figure}[!htb]
 \begin{center}
 \includegraphics[bb=0 0 1318 767, width=1.0\linewidth]{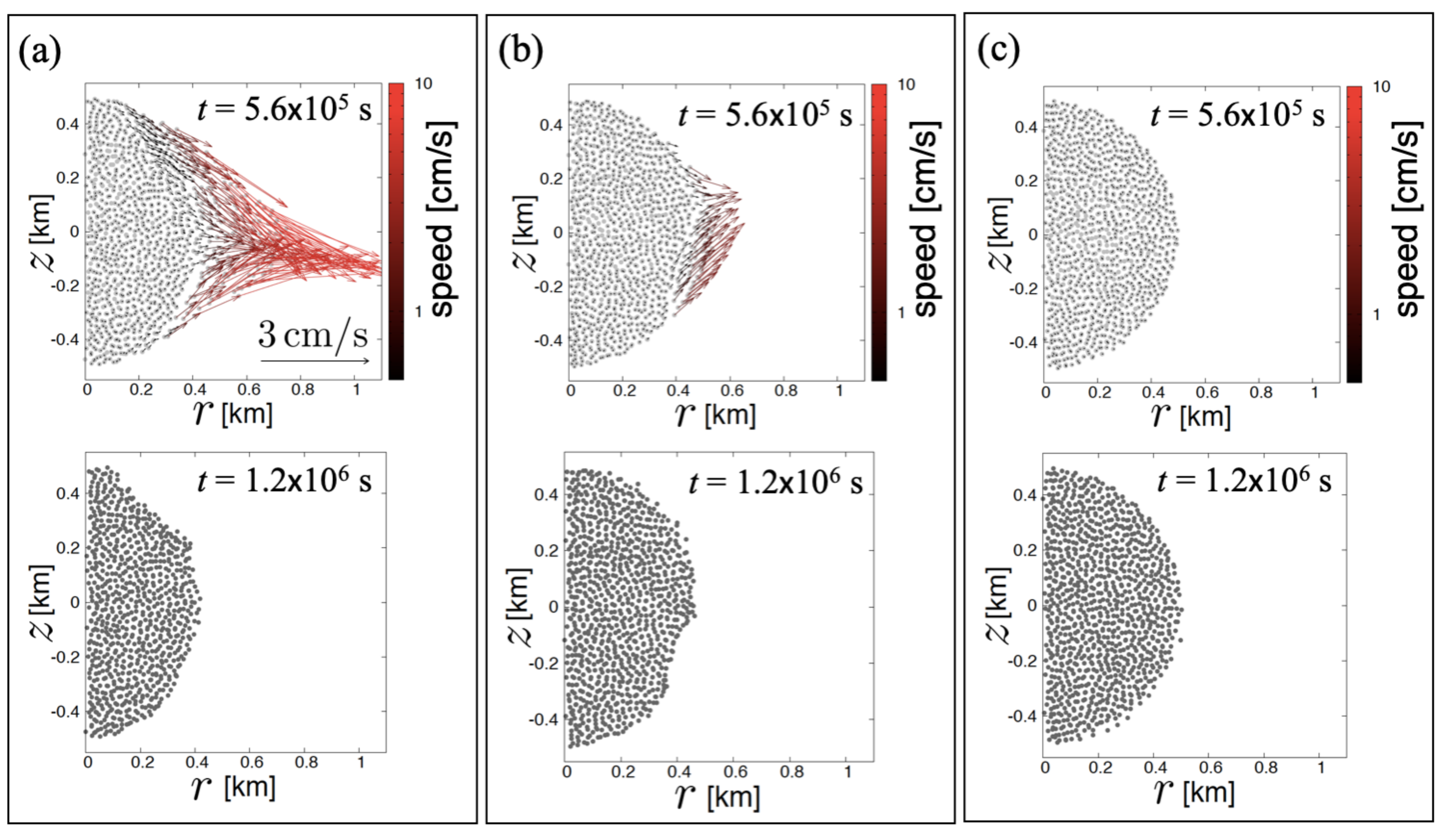}
 \caption{Meridional cross-sections of the main body obtained from the simulation with $\phi_{d}=80^{\circ}$ and $\beta = 0.05$. Panels (a), (b), and (c) show the cross-sections at the longitudes of $45^{\circ}$ E, $180^{\circ}$ E, and $270^{\circ}$ E, respectively. Here, the prime meridian of the main body passed the $+x$-direction at $t=0$. The upper panels show cross-sections of the main body with the velocity vectors at $t=5.6\times 10^{5}\,{\rm s}$, and the lower panels show cross-sections of the main body at $t=1.2\times 10^{6}\,{\rm s}$.}
 \label{2DPositionAndVelocityAtDeformation-gsr-R=500m-N=25000-0.05NominalSlowAccel} 
 \end{center}
\end{figure}

Figure \ref{2DPositionAndVelocityAtDeformation-gsr-R=500m-N=25000-0.05NominalSlowAccel} shows the cross-sections of the main body at different longitudes. The landslides occur at a longitude of $45^{\circ}$ E (Fig.\,\ref{2DPositionAndVelocityAtDeformation-gsr-R=500m-N=25000-0.05NominalSlowAccel}a), where the straight surface in the meridional cross-section appears in the northern hemisphere. However, only a small portion of mass is ejected from longitudes around $180^{\circ}$ E (Fig.\,\ref{2DPositionAndVelocityAtDeformation-gsr-R=500m-N=25000-0.05NominalSlowAccel}b), and no mass ejection occurs at longitudes around $270^{\circ}$ E (Fig.\,\ref{2DPositionAndVelocityAtDeformation-gsr-R=500m-N=25000-0.05NominalSlowAccel}c). The body's surface profiles in the cross-sections at longitudes around $180^{\circ}$ E and $270^{\circ}$ E maintain almost circular profiles.

Although the main body initially has an almost uniform density, small fluctuations in the SPH particle distribution cause slight locational variations in surface roughness and local slopes, which cause different stability levels against local landslides at different surface positions. In the case of the spinup with $\beta=1$, major landslides occur when $P = 3.03\,{\rm h}$ (Fig.\,\ref{pictures-gsr-R=500m-N=25000-nominalSlowAccel}k). Under such a fast rotation, landslides are induced everywhere by strong centrifugal forces regardless of local surface slopes. However, in the case of slower spinup with $\beta=0.05$, major landslides occur at an earlier time when $P = 3.08\,{\rm h}$ (Fig.\,\ref{pictures-gsr-R=500m-N=25000-0.05NominalSlowAccel}). Owing to the weaker centrifugal forces at that time, landslides seem to selectively occur on surfaces with locally steeper slopes, allowing surfaces with locally shallower slopes to remain mostly intact. This may result in localized landslides or a non-axisymmetric set of landslides, forming a non-axisymmetric shape. The angular momentum of the main body is lost due to the landslides in regions with locally steeper slopes, which leads to a decrease of the rotation rate. Thus, further landslides in the regions with locally shallower slopes do not occur.

\subsection{Spinup of a body with an equatorial ridge \label{Spinup-of-bodies-with-equatorial-ridges}}
\begin{figure}[!htb]
 \begin{center}
 \includegraphics[bb=0 0 1176 767, width=1.0\linewidth]{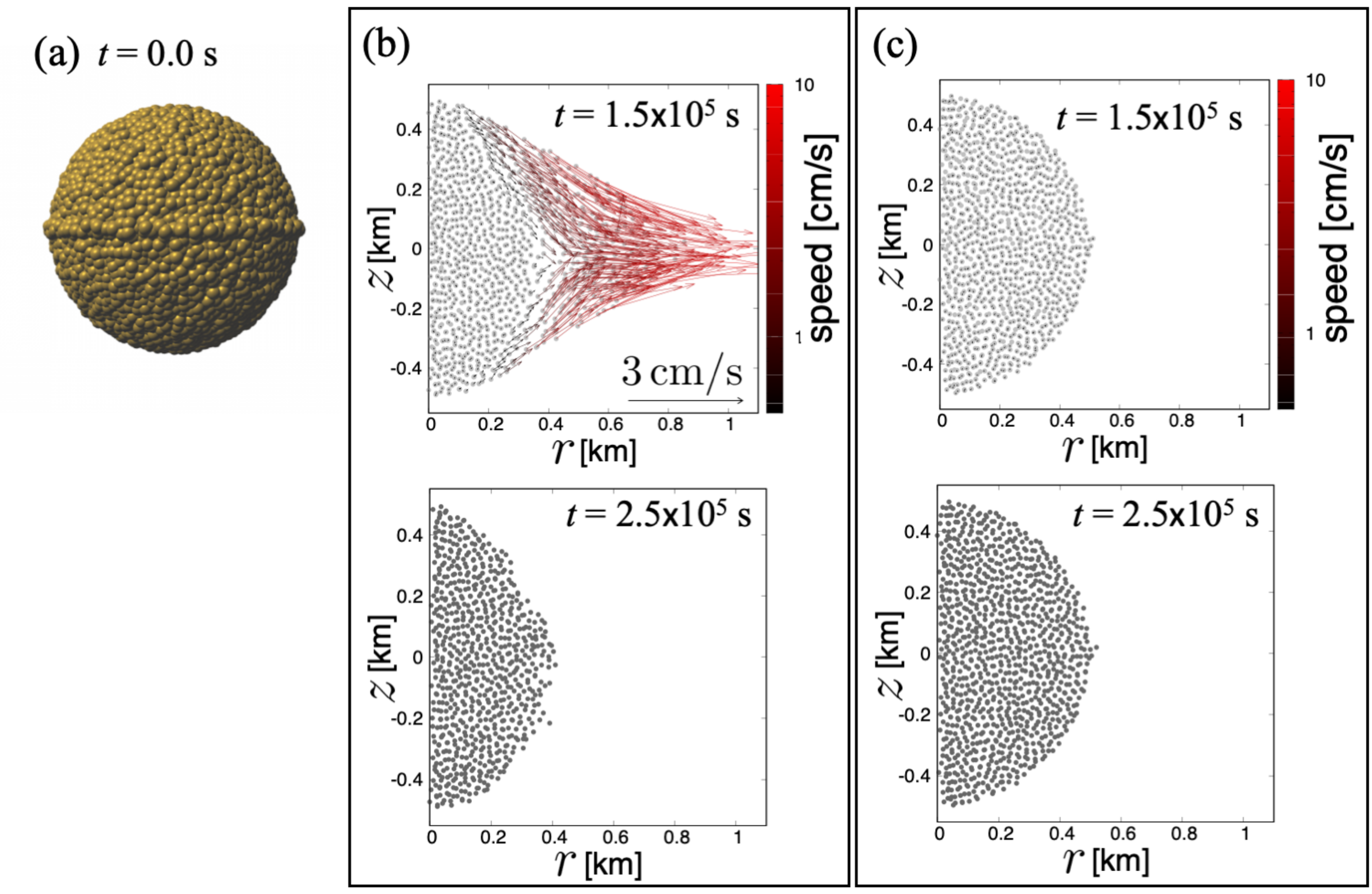}
 \caption{Initial shape and meridional cross-sections of a body with an equatorial ridge with $H_{{\rm ridge}} = 25\,{\rm m}$ and $\beta=1$. Panel (a) shows the initial shape of the body. Panels (b) and (c) show the cross-sections at longitudes of $90^{\circ}$ E and $270^{\circ}$ E, respectively. The upper panels show cross-sections of the main body with the velocity vectors at $t=1.5\times 10^{5}\,{\rm s}$, and the lower panels show the cross-sections of the main body at $t=2.5\times 10^{5}\,{\rm s}$.}
 \label{2DPositionAndVelocity-gsr-R=500m-with-25mRadiusRidge-N=25000-phid=80-nominalSlowAccel}
 \end{center}
\end{figure}

As shown in Section \ref{Spinup-with-beta=0.05}, even small fluctuations of SPH particle distribution cause non-axisymmetric distribution of landslides. Thus, we expect that a visible surface topographic high affects occurrence of landslides. We examined the spinup of a body with an equatorial ridge as an example of spinup of a body with surface topographic high. For simplicity, we put a half torus with a height of $H_{{\rm ridge}}$ on a spherical body to represent an equatorial ridge (see Fig.\,\ref{2DPositionAndVelocity-gsr-R=500m-with-25mRadiusRidge-N=25000-phid=80-nominalSlowAccel}a). We set the effective friction angle to $\phi_{d}=80^{\circ}$ and the initial rotation period to $4.0\,{\rm h}$.

We found that the simulation of spinup with $H_{{\rm ridge}} = 25\,{\rm m}$ and $\beta = 1$ results in a non-axisymmetric set of landslides. Figure \ref{2DPositionAndVelocity-gsr-R=500m-with-25mRadiusRidge-N=25000-phid=80-nominalSlowAccel}b and c show cross-sections of the body at two different longitudes. At a longitude of $90^{\circ}$ E (Fig.\,\ref{2DPositionAndVelocity-gsr-R=500m-with-25mRadiusRidge-N=25000-phid=80-nominalSlowAccel}b), landslides occur and a straight surface in this meridional cross section that resembles those seen in cross-sections of top shapes is produced. However, at the longitude of $270^{\circ}$ E (Fig.\,\ref{2DPositionAndVelocity-gsr-R=500m-with-25mRadiusRidge-N=25000-phid=80-nominalSlowAccel}c), no landslides occur and the resultant cross-section remains spherical. Thus, the spinup of a body with the surface topographic highs results in a non-axisymmetric set of landslides.

In contrast, the simulation with $H_{{\rm ridge}} = 25\,{\rm m}$ and faster spinup rate $\beta = 2$ results in an axisymmetric set of landslides, producing a top shape. In Sections \ref{spinup-with-beta=1} and \ref{Spinup-with-beta=0.05}, we see that the simulation with $H_{{\rm ridge}} = 0\,{\rm m}$ and $\beta = 1$ results in an axisymmetric set of landslides, but that with $H_{{\rm ridge}} = 0\,{\rm m}$ and $\beta = 0.05$ results in a non-axisymmetric set of landslides. These results suggest that critical spinup rates that are required for an axisymmetric set of landslides become slower for bodies with less surface roughness.
  
\section{Discussion \label{Discussion}}
Here, we discuss implications for the formation of real top-shaped asteroids in light of the results of our simulations using the SPH method.

\subsection{Cohesion and large effective friction angle \label{Cohesion-and-large-effective-friction-angle}}
Our simulations demonstrated that the effective friction angle $\phi_{d}$ of the constituent material mainly determines the deformation modes of the rotating bodies. We found that spinup of a spherical body with effective friction angles $\phi_{d} \leq 40^{\circ}$ induces quasi-static and internal deformation, whereas that with $\phi_{d} \geq 50^{\circ}$ results in dynamical deformation. In the dynamical regime, internal deformation occurs in the cases with $\phi_{d} \leq 60^{\circ}$, whereas surface landslides occur in $\phi_{d} \geq 70^{\circ}$. These transitions of the deformation modes seem to be irrespective of the spinup rate $\beta$ according to our simulations with $0.05 \leq \beta \leq 1$. However, for $\phi_{d} \geq 70^{\circ}$, axisymmetricity of the resultant shape depends on $\beta$; landslides occur almost axisymmetrically through fast spinup ($\beta=1$), while local landslides with a non-axisymmetric distribution occur through slow spinup ($\beta=0.05$).

A body with $\phi_{d} \leq 60^{\circ}$ deforms into an oblate spheroid through internal deformation, but never into a top shape that has conical surfaces extending from the equator to the northern and southern mid-latitudes. In contrast, a body with $\phi_{d} \geq 70^{\circ}$ can deform into an axisymmetric top shape through an axisymmetric set of landslides. Thus, we suggest that the formation of top shapes requires large effective friction angles of $\phi_{d} \geq 70^{\circ}$. As we discussed in Section \ref{Simulation-setup-and-analysis-of-results}, we interpreted $\phi_{d}$ used in our simulations as the effective friction angle that mimics an aspect of cohesion. Here, we show the validity of using large effective friction angles instead of introducing the cohesion.

Recall, the shear strength of granular material $Y_{d}$ is expressed as

\begin{equation}
  Y_{d} = \tan(\phi_{d,{\rm real}})p + c_{{\rm coh}},
  \label{shear-strength-of-granular-material}
\end{equation}

\noindent where $\phi_{d,{\rm real}}$ is the friction angle, $p$ is the confining pressure, and $c_{{\rm coh}}$ is the cohesion. Thus, the effective friction angle $\phi_{d}$, i.e., the arctangent value of the ratio of the shear strength to the confining pressure, is expressed as

\begin{equation}
  \phi_{d} = \tan^{-1}\Bigl( \frac{Y_{d}}{p} \Bigr) = \tan^{-1}\Bigl( \tan (\phi_{d,{\rm real}}) + \frac{c_{{\rm coh}}}{p} \Bigr).
  \label{effective-friction-angle}
\end{equation}

\begin{figure}[!htb]
 \begin{center}
 \includegraphics[bb=0 0 2163 540, width=1.0\linewidth]{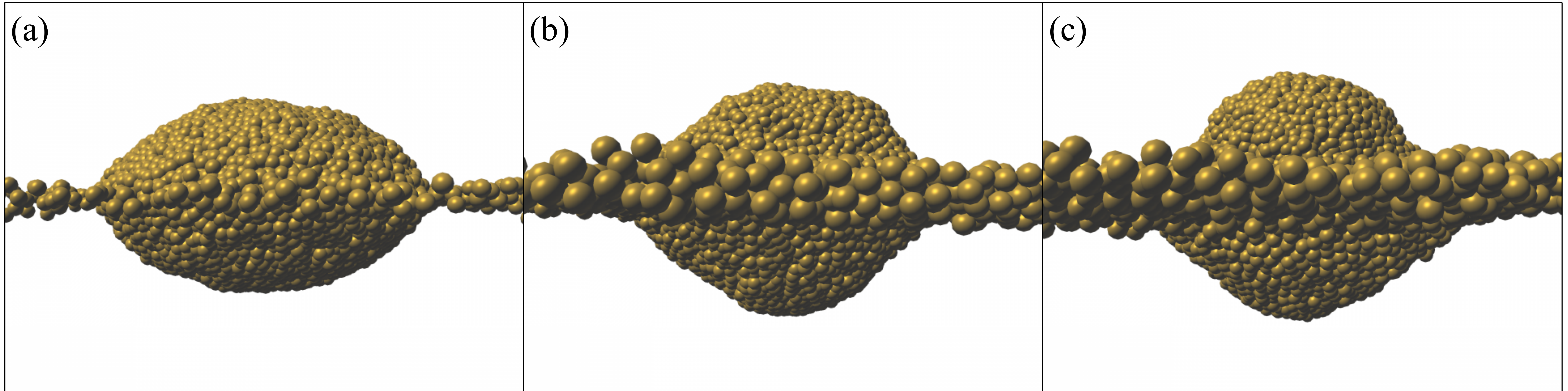}
 \caption{Snapshots of the bodies after their deformation obtained from three simulations with $\beta=1$ and the shear strength of Eq.\,(\ref{shear-strength-of-granular-material}). We set $\phi_{d,{\rm real}}=40^{\circ}$ and $c=30\,{\rm Pa}$(a), $50\,{\rm Pa}$(b), and $100\,{\rm Pa}$(c).}
 \label{pictures-gsr-with-shear-cohesions} 
 \end{center}
\end{figure}

We conducted numerical simulations with the same setup shown in Section \ref{spinup-with-beta=1} but with the shear strength of Eq.\,(\ref{shear-strength-of-granular-material}). We set $\phi_{d,{\rm real}}=40^{\circ}$ and $c_{{\rm coh}}=30,\,50$, and $100\,{\rm Pa}$, which, respectively, correspond to the effective friction angles of $\phi_{d}=55^{\circ},\,61^{\circ}$, and $71^{\circ}$ for the central pressure $p = (2/3)\pi G \rho_{0}^{2} R^{2} \approx 50\,{\rm Pa}$ of an asteroid with radius $R = 500\,{\rm m}$ and density $\rho_{0} = 1.19\,{\rm g/cm^{3}}$. Note that the central pressure is the maximum pressure in an asteroid and gives the minimum estimated value of effective friction angles. We found that the resultant shapes obtained from the simulations with $c_{{\rm coh}}=30,\,50$, and $100\,{\rm Pa}$ are an oblate spheroid (Fig.\,\ref{pictures-gsr-with-shear-cohesions}a), a slightly flattened top shape (Fig.\,\ref{pictures-gsr-with-shear-cohesions}b), and a top shape (Fig.\,\ref{pictures-gsr-with-shear-cohesions}c), respectively. The resultant shapes and axis ratios $c/a$ with $c_{{\rm coh}}=30, 50$, and $100\,{\rm Pa}$ are similar to those with effective friction angles of $\phi_{d}=60^{\circ}$ (Fig.\,\ref{pictures-gsr-R=500m-N=25000-nominalSlowAccel}j and Fig.\,\ref{2DPositionAndVelocityAtDeformation-gsr-R=500m-N=25000-nominalSlowAccel}b), $70^{\circ}$ (Fig.\,\ref{2DPositionAndVelocityAtDeformation-gsr-R=500m-N=25000-nominalSlowAccel}c), and $80^{\circ}$ (Fig.\,\ref{pictures-gsr-R=500m-N=25000-nominalSlowAccel}o and Fig.\,\ref{2DPositionAndVelocityAtDeformation-gsr-R=500m-N=25000-nominalSlowAccel}d), respectively. For example, $c/a$ of the body in Fig.\,\ref{pictures-gsr-R=500m-N=25000-nominalSlowAccel}o and Fig.\,\ref{pictures-gsr-with-shear-cohesions}c is 0.76 and 0.75, respectively. Thus, the simulations with the shear strength of Eq.\,(\ref{shear-strength-of-granular-material}) agree well with those with the shear strength of $Y_{d} = p \tan(\phi_{d})$ and corresponding effective friction angles $\phi_{d}$.

While we find the agreement, obviously Eq.\,(\ref{shear-strength-of-granular-material}) is not completely the same as $Y_{d}=p\tan (\phi_{d})$ even with corresponding effective friction angles $\phi_{d}$. For example, the confining pressures vary in post-yield flow of grain material, which may be minor effect for the spinup deformation of kilometer sized asteroids, and the shear strength with Eq.\,(\ref{shear-strength-of-granular-material}) somewhat deviates from that with $Y_{d}=p\tan (\phi_{d})$. The surface profile of the body shown in Fig.\,\ref{pictures-gsr-with-shear-cohesions}a is more bumpy than that with the effective friction angle $\phi_{d}=60^{\circ}$ (Fig.\,\ref{2DPositionAtFinal-gsr-R=500m-N=25000-nominalSlowAccel}), of which difference is quantitatively represented by the difference of $\delta_{V}$; $\delta_{V}$ for the former is 0.16 while that for the latter is 0.071. This may be caused by the difference of the constitutive laws. Moreover, we ignored the direct effect of cohesion, that is, tensile forces of cohesive materials, which may affect the results of our simulations. Investigation of the detailed effect of cohesion is planned for future work.

The cohesion of a subsurface layer of Ryugu $c_{{\rm coh}} = 140 - 670\,{\rm Pa}$ (\citealt{Arakawa-et-al2020}) results in the effective friction angle $\phi_{d} \approx 75 - 86^{\circ}$ with a typical confining pressure for a kilometer sized body $p \approx 50\,{\rm Pa}$ and a friction angle $\phi_{d,{\rm real}}=40^{\circ}$. Thus, large effective friction angles of $\phi_{d} \geq 70^{\circ}$ are plausible for $1\,{\rm km}$-sized asteroids. In contrast, a larger object has larger confining pressure and smaller effective friction angle. The central pressure of a $10\,{\rm km}$-sized asteroid is $p_{{\rm c}} \approx 5\,{\rm kPa}$, which leads to an effective friction angle $\phi_{d} \approx 44^{\circ}$ even with $c_{{\rm coh}} = 670\,{\rm Pa}$. Thus, our simulations imply that it would be difficult to form a top-shaped body with a diameter larger than $10\,{\rm km}$ through spinup, which is consistent with the fact that all top-shaped asteroids found so far are smaller than $10\,{\rm km}$.

\subsection{Formation of a top shape through fast spinup \label{Possible-formation-scenarios-of-axisymmertic-top-shapes}}
Our results show that top shapes are formed by the fast spinup ($\beta \geq 1$) of spherical bodies with effective friction angles $\geq 70^{\circ}$ (Fig.\,\ref{pictures-gsr-R=500m-N=25000-nominalSlowAccel}k-o). However, the slow spinup ($\beta=0.05$) of spherical bodies with $\geq 70^{\circ}$ induces a non-axisymmetric set of landslides and results in non-axisymmetric shapes (Fig.\,\ref{pictures-gsr-R=500m-N=25000-0.05NominalSlowAccel}). Moreover, surface roughness or topography suppresses top-shape formation. The fast spinup ($\beta = 1$) of a body with a $25\,{\rm m}$ high equatorial ridge results in a non-axisymmetric shape (Fig.\,\ref{2DPositionAndVelocity-gsr-R=500m-with-25mRadiusRidge-N=25000-phid=80-nominalSlowAccel}).

We discuss the formation of axisymmetric top shapes through YORP spinup (Section \ref{YORP-Effect}). We also discuss fast spinup through the reaccumulation of fragments after catastrophic collisions (Section \ref{Reaccumulation-of-fragments}), although applying our spinup simulations to spinup through the accretion requires the assumption that mass increase and shape changes are negligible through the accretion, which is discussed in Section \ref{Reaccumulation-of-fragments}. In addition, we discuss possibilities of other spinup mechanisms (Section \ref{Other-mechanisms-with-short-spinup-timescales}). Here, instead of spinup rate $\beta$, we use spinup timescale $t_{{\rm su}}$, which is defined as the elapsed time to spin up the body from rotation period $P=3.5\,{\rm h}$ to $P=3.0\,{\rm h}$: $t_{{\rm su}} \approx 9.3 \times 10^{4} \beta^{-1} \,{\rm s}$.

\subsubsection{YORP effect \label{YORP-Effect}}
The YORP effect for a kilometer-sized asteroid causes spinup with a timescale $t_{{\rm su}} \gtrsim 100\,{\rm kyr} \approx 3\times 10^{12}\,{\rm s}$ (e.g.,\,\citealt{Kaasalainen-et-al2007, Hergenrother-et-al2019}). We showed that a non-axisymmetric set of landslides tends to occur under slower spinups even in the case with $\beta = 0.05$ or $t_{{\rm su}} \approx 1.9\times 10^{6}\,{\rm s}$. This suggests that a non-axisymmetric set of landslides and the formation of a non-axisymmetric shape are likely results from the YORP spinup of an asteroid.

We showed that a body with less surface roughness experiences an axisymmetric set of landslides even for slower spinup. The kilometer-sized spherical bodies with 25,000 SPH particles initially set in our simulations are estimated to have a surface roughness of $\sim 10\,{\rm m}$ due to discretization of the bodies using SPH particles. The numerical roughness is larger than typical surface roughness of asteroids (e.g.,\,\citealt{Sugita-et-al2019}), and thus the spinup timescale of asteroids required for an axisymmetric set of landslides may be longer than what we obtained in the simulations. We extrapolated the results obtained in Sections \ref{Spinup-with-beta=0.05} and \ref{Spinup-of-bodies-with-equatorial-ridges} through a power-law fitting and found that an axisymmetric set of landslides for bodies with $1\,{\rm cm}$ surface roughness requires $t_{{\rm su}}\sim 10^{7}\,{\rm s}$; the required spinup timescale is still much shorter than the timescale of the YORP effect.

However, we do not rule out the possibility that the YORP spinup will produce a top-shaped asteroid. The YORP effect may spin up an asteroid even after a major landslide event if deformation and topographical changes caused by the major landslides insignificantly alter the direction of the net YORP torque exerted on the body, which eventually causes further landslides. The place where a local landslide has occurred has a locally conical surface and has been stabilized against further landslides. Thus, the subsequent landslides due to further YORP spinup may occur at the longitudes where significant local landslides have not occurred yet, and eventually multiple occurrences of local landslides at different longitudes would produce an almost axisymmetric top shape. Since we only investigated a shape formed until a single major landslide event, we do not rule out the possibility of the formation of top shapes through multiple cycles of the YORP spinup and local landslides.

\subsubsection{Reaccumulation of fragments \label{Reaccumulation-of-fragments}}
The reaccumulation phase after catastrophic disruption of the parent body of a rubble-pile asteroid by a collision is yet another practical situation for fast spinup of the asteroid through accretion (\citealt{Sugiura-et-al2019PSS,Michel-et-al2020}). Note that \cite{Michel-et-al2020} focus on the possibility of top-shape formation through deformation due to the accretion process, but we focus on top-shape formation through spinup due to accretion. The catastrophic collision produces a sheet-like structure composed of fragments, and dense parts of the sheet gravitationally collapse into small cores of remnants. Then, the remaining fragments accumulate on the remnants to supply angular momentum. Note that the supplied angular momentum originates from the small part of the sheet with not random but specific velocity field around it, so that the accretion of the fragments typically causes continuous spinup of the remnant. Moreover, the timescale of the reaccumulation is $\sim (G \rho_{0})^{-1/2} \approx 4 \times 10^{3}\,{\rm s}$, which is shorter than the spinup timescale $t_{{\rm su}}$ required for the formation of an axisymmetric top shape $t_{{\rm su}} \sim 10^{5}\,{\rm s}$. The fast spinup caused through the reaccumulation may form a top shape.

Accretion of fragments supplies not only angular momentum but also mass onto the core. Moreover, accretion may induce surface mass flow, which may affect how subsequent landslides occur. These effects were not represented in our simulations; we just accelerated the spin of the initial body. However, the mass of fragments required for spinup may be negligible compared to the core mass. We estimated angular momentum increase due to the accretion as $M_{\rm f} R v_{{\rm esc}}$, where $M_{{\rm f}}$ is the total mass of fragments accreting on the core, $R = 500\,{\rm m}$ is the radius of the core, and $v_{{\rm esc}} \approx 40\,{\rm cm/s}$ is the escape speed from the core. Note that $M_{\rm f} R v_{{\rm esc}}$ roughly explains angular momentum increase through accretion in the simulations of the catastrophic collisions (\citealt{Sugiura-et-al2019PSS}). The amount of fragments required to accelerate the core's rotation period from $3.5\,{\rm h}$ to $3.0\,{\rm h}$ is estimated to be only 4\% of the core mass. Thus, our simulations of the spinup possibly mimic the spinup due to the reaccumulation of fragments. The possibility remains to be studied in future works.

\subsubsection{Other mechanisms with short spinup timescales \label{Other-mechanisms-with-short-spinup-timescales}}
Acceleration rates until the rotation periods of bodies reach the critical rotation period $\sim 3\,{\rm h}$ are not related to how landslides occur. Thus, the following formation procedure of axisymmetric top shapes is possible: the YORP effect accelerates the rotation periods of bodies up to $\sim 3\,{\rm h}$ and then other mechanisms with short spinup timescales further accelerate the bodies and cause an axisymmetric set of landslides. In this case, small amounts of angular momentum supply due to the fast spinup mechanisms are sufficient to cause an axisymmetric set of landslides.

A non-destructive impact of a small asteroid supplies angular momentum and may cause fast spinup. A prograde impact (i.e., angular momentum vector provided by an impactor is parallel to that of a rotating body) tends to cause tentative spinup while the impactor contacts with the rotating body. Since the impact may erode the body and take away angular momentum to some extent, the process is not straightforward and thus this phenomena should be investigated via simulations in future works.

An asteroid passing around the Roche radius of a planet acquires angular momentum via tidal torque (\citealt{Hyodo-et-al2016}), while an encounter much closer than the Roche radius results in elongation and disruption of the asteroid (e.g.,\,\citealt{Walsh-and-Richardson2006}). A moderate encounter may result in the spinup of an asteroid with keeping its shape. All the top-shaped asteroids found so far are near-Earth asteroids, so that a close encounter with the Earth is one possible spinup mechanism for the formation of top-shaped asteroids.

\subsection{Subsequent evolution of ejected materials via landslides}
Our simulations showed that an axisymmetric set of landslides that produces a top-shaped asteroid ejects fragments with mass $\sim 0.1 M_{{\rm b}}$ (Section \ref{spinup-with-beta=1}), where $M_{{\rm b}}$ is the mass of the central body. The ejected fragments produce satellites with $\sim 10 \%$ of the total fragments (\citealt{Hyodo-et-al2015}). The mass of the satellites $\sim 1\%$ of host asteroids is consistent with those of satellites around observed top-shaped asteroids (\citealt{Ostro-et-al2006, Brozovic-et-al2011, Becker-et-al2015}). However, about half of observed top-shaped asteroids do not have satellites (e.g.,\,Ryugu or Bennu). This suggests that there is a mechanism to remove satellites from the top-shaped asteroids within their lifetimes.

An orbit of a satellite is mainly altered by tidal interaction with a host asteroid (e.g.,\,\citealt{Goldreich-and-Sari2009}), the binary YORP (BYORP) effect (e.g.,\,\citealt{Cuk-and-Burns2005, McMahon-and-Scheeres2010a, McMahon-and-Scheeres2010b}), and a close encounter with a planet (e.g.,\,\citealt{Walsh-and-Richardson2008}). The YORP effect is probably the dominant satellite-evolution mechanism for a binary near-Earth asteroids with binary separation $\sim$ the Roche radius of a host asteroid, although the orbital evolution through each mechanism highly depends on various physical parameters. The BYORP effect is an analog of the YORP effect for a binary system with a satellite spinning synchronously with its orbital motion, and torque induced by the absorption of sunlight by and the thermal re-emission from the irregularly shaped satellite changes the orbit of the satellite. The timescale of the secular evolution of the semimajor axis $a$ due to the BYORP effect is estimated as (\citealt{McMahon-and-Scheeres2010a, Walsh-and-Jacobson2015})

\begin{align}
\tau_{{\rm B}} \equiv \frac{a}{|\dot{a}_{{\rm BYORP}}|} & = \frac{2 \pi \rho_{0} \omega_{d} R_{h}^{5/2} (M_{s}/M_{h})^{1/3}}{3 H_{\odot} B_{s} a^{1/2} \sqrt{1 + (M_{s}/M_{h})}} \nonumber \\ & = 5.5 \times 10^{4} \Bigl( \frac{M_{s}/M_{h}}{0.01} \Bigr)^{1/3} \Bigl( \frac{R_{h}}{500\,{\rm m}} \Bigr)^{2} \Bigl( \frac{B_{s}}{0.01} \Bigr)^{-1} \Bigl( \frac{a}{R_{h}} \Bigr)^{-1/2}\,{\rm yr}, \label{BYORP-evolution-timescale}
\end{align}

\noindent where $\rho_{0}$ is the bulk density of the binary asteroids, $\omega_{d}=\sqrt{(4/3)\pi G \rho_{0}}$ is the critical spin frequency, $R_{h}$ is the radius of the host asteroid, $M_{s}/M_{h}$ is the mass ratio of the satellite and the host asteroid, the heliocentric orbit factor $H_{\odot}$ is $4\times 10^{-5}\,{\rm g \, cm^{-1} \, s^{-2}}$ at $1\,{\rm au}$ (\citealt{Scheeres2007}), and the dimensionless parameter $B_{s}$ that depends on the satellite's shape is estimated to be $\sim 0.01$ (\citealt{McMahon-and-Scheeres2010a, Walsh-and-Jacobson2015}). If the BYORP effect causes the outward migration of the satellite, the satellite is ejected from the binary system within $\sim 10^{5}\,{\rm yr}$, which is much shorter than the dynamical lifetime of near-Earth asteroids $\sim 10\,{\rm Myr}$ (\citealt{Gladman-et-al2000}). Thus the BYORP effect can eject the satellite during its stay in the near-Earth space, which may explain the lack of satellites for several top-shaped asteroids. 

\section{Summary \label{Summary}}
Rubble-pile asteroids Ryugu and Bennu visited by the spacecraft Hayabusa2 and OSIRIS-REx, respectively, have a similar spinning-top shape with an axisymmetric equatorial ridge. Many top-shaped asteroids rotate near their break-up periods of $\sim 3$ hours. The axisymmetricity and present-day high spin rates suggest that these top-shaped asteroids were formed through some spinup processes. However, how a rubble-pile body with a specific friction angle deforms under various spinup rates is still under debate. 

Given the radius and bulk density, we numerically simulated the spinup of a spherical rubble-pile body with different effective friction angles $\phi_{d}$ using the SPH method for granular material. The high $\phi_{d}$ mimics the effect of shear cohesion, and we confirmed that the simulation results with $\phi_{d}$ are similar to those with the real cohesive strength law (Eq.\,(\ref{shear-strength-of-granular-material})) with corresponding values of cohesion. The non-dimensional acceleration rate of the rotation was defined as $\beta = \dot{\omega}_{0} / \dot{\omega}_{n}$, where a normalizing acceleration rate $\dot{\omega}_{n} = 8.954\times 10^{-10}\,{\rm rad/s^{2}}$ corresponds to a decrease in rotation period from $3.5\,{\rm h}$ to $3.0\,{\rm h}$ within $9.3\times 10^{4}\,{\rm s}$ ($\approx 8.0$ rotations). We also defined the spinup timescale as $t_{{\rm su}} \approx 9.3 \times 10^{4}\beta^{-1}\,{\rm s}$.

We mainly investigated spinups of a spherical body with an acceleration rate $\beta=1$ ($t_{{\rm su}} \approx 9.3\times 10^{4}\,{\rm s}$) and effective friction angles $\phi_{d}=20 - 80^{\circ}$. Contrary to the fluid cases ($\phi_{d}=0^{\circ}$), our simulations show that a moderate effective friction angle ($\phi_{d} \geq 30^{\circ}$) will suppress the deformation into a triaxial ellipsoid. In the cases with $\phi_{d} \leq 40^{\circ}$, quasi-static and internal deformation occurred, forming oblate spheroids (Fig.\,\ref{pictures-gsr-R=500m-N=25000-nominalSlowAccel}a-e). The simulations with $50^{\circ} \leq \phi_{d} \leq 60^{\circ}$ resulted in dynamical internal deformation and formed oblate spheroids (Fig.\,\ref{pictures-gsr-R=500m-N=25000-nominalSlowAccel}f-j, Fig.\,\ref{2DPositionAndVelocityAtDeformation-gsr-R=500m-N=25000-nominalSlowAccel}a, b, and Fig.\,\ref{2DPositionAtFinal-gsr-R=500m-N=25000-nominalSlowAccel}a, b), indicating that internal deformation will not lead to top-shape formation. In contrast, spinup of spherical bodies with $\phi_{d} \geq 70^{\circ}$ induced an axisymmetric set of surface landslides, producing axisymmetric top shapes (Fig.\,\ref{pictures-gsr-R=500m-N=25000-nominalSlowAccel}k-o, Fig.\,\ref{2DPositionAndVelocityAtDeformation-gsr-R=500m-N=25000-nominalSlowAccel}c, d, and Fig.\,\ref{2DPositionAtFinal-gsr-R=500m-N=25000-nominalSlowAccel}c, d).

We also investigated spinup with a slower acceleration rate $\beta=0.05$ ($t_{{\rm su}} \approx 1.9\times 10^{6}\,{\rm s}$) for comparison. We found that the change of $\beta$ probably does not affect the principal deformation modes found in the nominal case ($\beta=1$): quasi-static and internal deformation for $\phi_{d} \leq 40^{\circ}$; dynamical and internal deformation for $50^{\circ} \leq \phi_{d} \leq 60^{\circ}$; and a set of surface landslides for $\phi_{d} \geq 70^{\circ}$. However, the slower $\beta$ affects axisymmetricity of the distribution of landslides for $\phi_{d} \geq 70^{\circ}$. The simulation with $\phi_{d} = 80^{\circ}$ and $\beta=0.05$ ($t_{{\rm su}} \approx 1.9\times 10^{6}\,{\rm s}$) showed that a non-axisymmetric shape is formed through a set of local landslides. 

We showed that the formation of axisymmetric top shapes through spinup requires an effective friction angle $\phi_{d} \geq 70^{\circ}$, which is much larger than those of cohesionless granular materials. However, estimated small values of the cohesion for kilometer-sized asteroids ($c_{{\rm coh}} \sim 100\,{\rm Pa}$) make effective friction angles larger than $70^{\circ}$. Thus, the cohesion of asteroids plays an important role for the formation of axisymmetric top shapes.

In addition, our simulations showed that the formation of axisymmetric top shapes prefers fast spinup with timescales shorter than $\sim 10^{6}\,{\rm s}$. The timescale of the YORP spinup ($\gtrsim 100\,{\rm kyr}$) is much longer than the spinup timescales required for the formation of axisymmetric top shapes, suggesting a difficulty to produce axisymmetric top shapes via deformation through YORP spinup itself. On the other hand, catastrophic collisions of parent asteroids produce many small remnants and accretion of fragments onto these remnants may spins them up within the reaccumulation timescale $\sim 4 \times 10^{3}\,{\rm s}$. Thus, catastrophic collisions are one possible origin of axisymmetric top shapes. Besides reaccumulation of fragments after catastrophic collisions, coalescence of fragments that are ejected through the YORP spinup, a non-destructive and prograde impact of an asteroid, and tidal torque caused by a close encounter with a planet may spin up asteroids with short timescales and may lead to the formation of top shapes. Our simulations showed that the formation of top-shaped asteroids is accompanied by surface mass ejection, which leads to the formation of satellites. About half of observed top-shaped asteroids do not have satellite, and this can be explained by the ejection of satellites such as outward migration of satellites' orbits due to the BYORP effect.

\section*{Acknowledgements}
K.S. acknowledges the financial support of JSPS KAKENHI Grant (JP20K14536, JP20J01165) and the financial support as JSPS Research Fellow. K.S. and H.G. were supported by MEXT KAKENHI Grant No. JP17H06457. H.K acknowledges the financial support of JSPS KAKENHI Grant (JP17H01103, JP17K05632, JP17H01105, JP18H05438, JP18H05436, JP20H04612). S.W acknowledges KAKENHI support from JSPS (JP17H06459, JP19H01951). R.H. acknowledges the financial support of JSPS Grants-in-Aid (JP17J01269, 18K13600). Numerical simulations in this work were carried out on the Cray XC50 supercomputer at the Center for Computational Astrophysics, National Astronomical Observatory of Japan.

\section*{Appendix A. Dependence of resultant shapes on the size, the sound speed, and equations of state}

\begin{figure}[!htb]
 \begin{center}
 \includegraphics[bb=0 0 2163 540, width=1.0\linewidth]{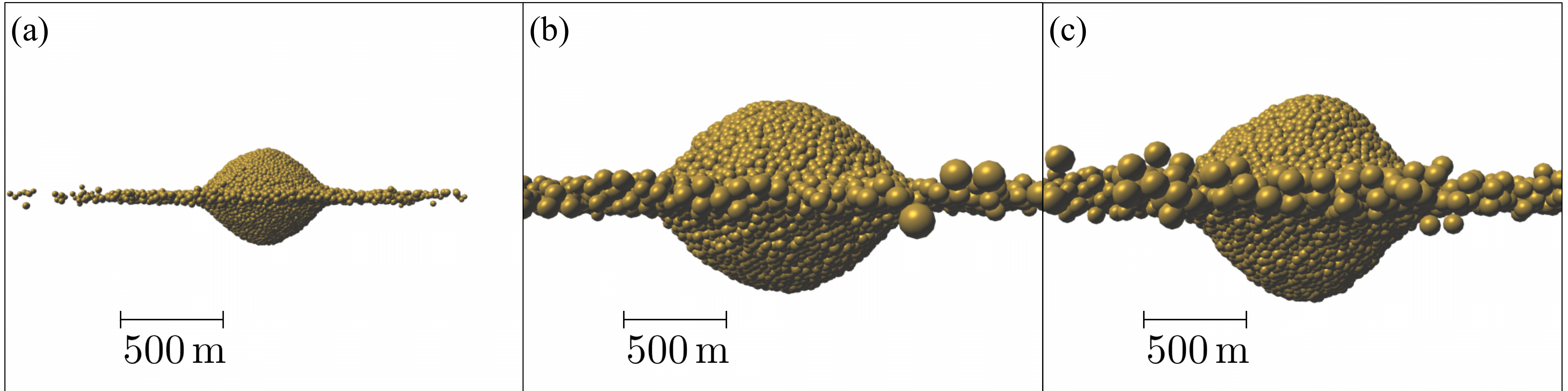}
 \caption{Snapshots of the bodies after their deformation obtained from three simulations with effective friction angle $\phi_{d}=80^{\circ}$ and acceleration rate $\beta=1$. Panel (a) shows the simulation with the body with radius $R=250\,{\rm m}$, panel (b) shows that with the sound speed $C_{s}=300\,{\rm m/s}$, and panel (c) shows that with the Tillotson equation of state.}
 \label{pictures-gsr-differentEoS-and-radius} 
 \end{center}
\end{figure}

Here, we show that the results of spinup simulations are almost independent of the radius of bodies, the value of the sound speed, and the choice of equations of state. The radius, the sound speed, and the equation of state in nominal simulations are $500\,{\rm m}$, $100\,{\rm m/s}$, and Eq.\,(\ref{elastic-EoS}), respectively. We investigated the deformation of the spherical bodies with the effective friction angle $\phi_{d}=80^{\circ}$ through spinup with the acceleration rate $\beta=1$. We used the same conditions as those in the simulation of Fig.\,\ref{pictures-gsr-R=500m-N=25000-nominalSlowAccel}k-o, but we used a different radius of the body $R=250\,{\rm m}$ (Fig.\,\ref{pictures-gsr-differentEoS-and-radius}a), a different value of the sound speed $C_{s}=300\,{\rm m/s}$ (Fig.\,\ref{pictures-gsr-differentEoS-and-radius}b), and the Tillotson equation of state (Fig.\,\ref{pictures-gsr-differentEoS-and-radius}c). For the simulation with the Tillotson equation of state (Fig.\,\ref{pictures-gsr-differentEoS-and-radius}c), we used the same parameter for basaltic material described in \cite{Benz-and-Asphaug1999} except for the density $\rho=1.19\,{\rm g/cm^{3}}$ and the bulk modulus $A=B=1.19\times 10^{8}\,{\rm dyne/cm^{2}}$, which leads to the sound speed $C_{s}=100\,{\rm m/s}$.

In these three simulations, we observed similar axisymmetric sets of landslides to those shown in Fig.\,\ref{pictures-gsr-R=500m-N=25000-nominalSlowAccel}k-o. Although the axis ratios of the bodies resulting from these simulations are slightly different, they have top shapes with cone-like surfaces similar to the body shown in Fig.\,\ref{pictures-gsr-R=500m-N=25000-nominalSlowAccel}o. Thus, the dependence on the size, the sound speed, and the choice of equations of state are minor for our simulations with the given effective friction angle $\phi_{d}$. 

\bibliography{mybibfile}

\end{document}